\documentclass[aps,prd,twocolumn,superscriptaddress]{revtex4-1}

\usepackage{graphicx}

\begin{document}


\title{Interferometer design of the KAGRA gravitational wave detector}


\author{Yoichi Aso}
\email[]{aso@granite.phys.s.u-tokyo.ac.jp}

\author{Yuta Michimura}
\email[]{michimura@granite.phys.s.u-tokyo.ac.jp}
\affiliation{Department of Physics, University of Tokyo, 7-3-1 Hongo, Bunkyo-ku, Tokyo, 113-0033, JAPAN}

\author{Kentaro Somiya}
\email[]{somiya@phys.titech.ac.jp}
\affiliation{Department of Physics, Tokyo Institute of Technology, 2-12-1, Ookayama, Meguro-ku, Tokyo, 152-8550, JAPAN}

\author{Masaki Ando}
\affiliation{Department of Physics, University of Tokyo, 7-3-1 Hongo, Bunkyo-ku, Tokyo, 113-0033, JAPAN}
\affiliation{National Astronomical Observatory of Japan, 2-21-1 Osawa, Mitaka, Tokyo , 181-8588, JAPAN}

\author{Osamu Miyakawa}
\author{Takanori Sekiguchi}
\affiliation{Institute for Cosmic Ray Research, University of Tokyo, 5-1-5 Kashiwa-no-Ha, Kashiwa City, Chiba, 277-8582, JAPAN}

\author{Daisuke Tatsumi}
\affiliation{National Astronomical Observatory of Japan, 2-21-1 Osawa, Mitaka, Tokyo , 181-8588, JAPAN}

\author{Hiroaki Yamamoto}
\affiliation{LIGO Laboratory, California Institute of Technology, Pasadena, California, 91125, USA}

\collaboration{The KAGRA Collaboration}
\noaffiliation

\date{\today}

\begin{abstract}
  KAGRA is a cryogenic interferometric gravitational wave detector
  being constructed at the underground site of Kamioka mine in Gifu
  prefecture, Japan. We performed an optimization of the interferomter
  design, to achieve the best sensitivity and a stable operation, with
  boundary conditions of classical noises and under various practical
  constraints, such as the size of the tunnel or the mirror cooling
  capacity.  Length and alignment sensing schemes for the robust
  control of the interferometer are developed. In this paper, we
  describe the detailed design of the KAGRA interferometer 
  as well as the reasoning behind design choices.
\end{abstract}

\pacs{95.55.Ym, 42.60.Da}

\maketitle

\section{Introduction}
Direct detection of gravitational waves from astronomical sources will
not only be a powerful way to test gravity theories under strong
gravitational fields, but also an intrinsically new way to observe the
universe\,\cite{sathyaprakash_physics_2009}. Such observations will
provide us with unique information not available with conventional
astronomical observations using electromagnetic waves. Currently, one
of the most promising way to detect gravitational waves is to use
large laser interferometers. Several large-scale interferometric
gravitational wave detectors were built and successfully operated to
prove the feasibility of such
detectors\,\cite{hough_long_2011}. However, those first generation
detectors were still not sensitive enough to detect gravitational
waves unless there is an extremely lucky event, such as a nearby
neutron star merger. There are several next generation interferometric
gravitational wave detectors being built around the
world\cite{hough_long_2011}. These detectors generally aim at
improving the sensitivity by ten-fold from the first generation
detectors to make the regular detection a reality.

KAGRA is Japanese next-generation gravitational wave detector, now
under construction at the underground site of currently disused
Kamioka-mine, in Gifu prefecture, Japan. KAGRA has two outstanding
features: cryogenic mirrors made of mono-crystalline sapphire to
reduce thermal noises and a seismically quiet and stable
environment of the underground site. The construction of KAGRA started
in 2010 and it is planed to start the operation of the detector at its
full configuration in 2017.

The development of KAGRA is performed in two phases. The initial KAGRA,
or iKAGRA, is the fist phase of the operation with a simple Fabry-Perot
Michelson interferometer configuration. The main purpose of iKAGRA is to quickly
identify facility related or any other problems at an early stage of
construction, thus allowing more time to address those potential
issues. The final configuration of KAGRA is called the baseline KAGRA,
or bKAGRA. In this paper, we focus on the bKAGRA interferometer and
explain its design and the reasoning behind the parameter choices and so on.

The paper is structured as follows. First, we
give a brief overview of the KAGRA interferometer configuration and
set the terminology for the later discussion (Section
\ref{sec:overv-kagra-interf}). Then, we briefly review the noise
sources of KAGRA with non-quantum origins: namely, seismic noise
and thermal noises (Section \ref{Section: Non-q noise}). We treat
these noises as boundary conditions for optimizing the quantum
noises (shot noise and radiation pressure noise) in section
\ref{Section: QN Opt}. This process basically determines the
reflectivities of the interferometer mirrors. Then we proceed to
consider how to control those mirrors and lock the interferometer at
the optimal operation point in section \ref{Section: RFSB}. This boils
down to selecting the macroscopic lengths of the recycling cavities
and the Michelson asymmetry to realize the optimal resonant conditions
for the RF sidebands used to extract error signals for the
interferometer control.
In section
\ref{Section: Spatial modes}, we consider the spatial mode properties of the
interferometer, especially in terms of the ability to reject unwanted
higher-order modes. This determines the radii of curvature (ROCs) of
the mirrors. In section \ref{Section: Alignment}, we examine whether reasonable
alignment information of the interferometer can be extracted with the
selected interferometer parameters. Finally, we give conclusions in section
\ref{Section: Conclusion}.

\section{Overview of the KAGRA interferometer and the terminology}
\label{sec:overv-kagra-interf}
Before going into the details of the interferometer design,
we first give a brief overview of the interferometer configuration of
KAGRA and explain the terminology used throughout the rest of this
paper.

\subsection{Interferometer configuration}
\label{sec:interf-conf}
The schematic view of the KAGRA interferometer is shown in Figure
\ref{Figure: IFO Schematic}.  The laser beam is first passed through a
three-mirror optical cavity called mode cleaner (MC) to clean the
spatial mode of the incident beam. After the MC is a main
interferometer, which consists of 4 cryogenic mirrors and 7 auxiliary
mirrors at room temperature. We have two 3km-long Fabry-Perot
cavities, called arm cavities, formed by input test masses (ITMs) and
end test masses (ETMs). These test masses are cooled down to around
20\,K to reduce thermal noises. The two arm cavities are combined by a
beam splitter (BS) and the interference condition on the BS is held
such that all the light comes back to the direction of a mirror called
PR3. A power recycling mirror (PRM) and the two ITMs form a power
recycling cavity (PRC). Similarly, a signal recycling mirror (SRM)
forms a signal recycling cavity (SRC) together with the ITMs. This
interferometer configuration with two recycling cavities is called
dual-recycling. In particular, we keep the SRC length to be resonant
for the carrier light, which is called the resonant sideband
extraction (RSE) scheme.

The PRC and the SRC are folded in Z-shapes by two additional mirrors
each for improving the spatial mode stability as explained in
\ref{Section: Spatial modes}. At the downstream of the SRM, there is
an output mode cleaner (OMC) used to remove unwanted higher-order
spatial modes from the output beam.

\begin{figure}
\includegraphics[width=8cm]{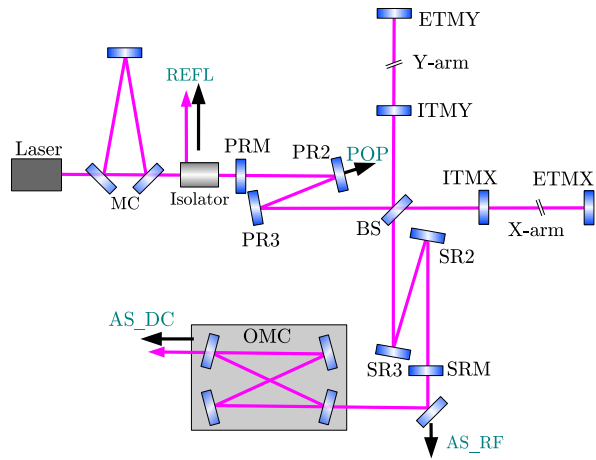}
\caption{Schematic of the KAGRA interferometer. Names of the mirrors as
  well as the signal detection ports are shown.\label{Figure: IFO
    Schematic}}
\end{figure}

\subsection{Length degrees of freedom}
\label{Section: DOF} From the point of view of interference and
resonance of the light, there are five length degrees of freedom
(DOFs) in our interferometer. The names of the DOFs are summarized in
Table \ref{Table: LDOF}. All the DOFs are represented as linear
combinations of the motions of the mirrors. The most important DOF is
the differential length change of the arm cavities, called DARM. It
contains gravitational wave information.  CARM is the common
change of the arm cavity lengths. MICH, which is short for Michelson,
is the differential change of the distances between the BS and the two
ITMs. PRCL and SRCL are the lengths of the PRC and the SRC respectively.
Since DARM is the most important DOF, other 4 DOFs are often called
auxiliary DOFs.

\begin{table}
\caption{Length degrees of freedom of the KAGRA
 interferometer.\label{Table: LDOF}}
\begin{ruledtabular}
\begin{tabular}{l|l}
DARM\hspace{2mm} &Differential length change of the arm cavities. \\
CARM &Common length change of the arm cavities.\\
MICH &Michelson degree of freedom. \\
PRCL &Power recycling cavity length. \\
SRCL &Signal recycling cavity length. \\
\end{tabular}
\end{ruledtabular}
\end{table}

\subsection{Detection ports}
\label{sec:sign-detect-ports}
Laser beams coming out of the interferometer are detected at various
places for extracting the interferometer information. The names of
those detection ports are also given in figure \ref{Figure: IFO
  Schematic}. The reflection port (REFL) is located at the reflection
output of a Faraday isolator, which reflects the light coming back
from the interferometer. The light coming out of the SRC is led to two
anti-symmetric (AS) ports. The beam picked off before the OMC goes to
the AS\_RF port. The transmission of the OMC is called the AS\_DC
port. The POP (Pick-Off-in-the-PRC) port is the transmission of the
PR2 mirror.

\section{Non-quantum noises of KAGRA}
\label{Section: Non-q noise}

\begin{figure*}[tbp]
  \begin{minipage}{8cm}
\includegraphics[width=8cm]{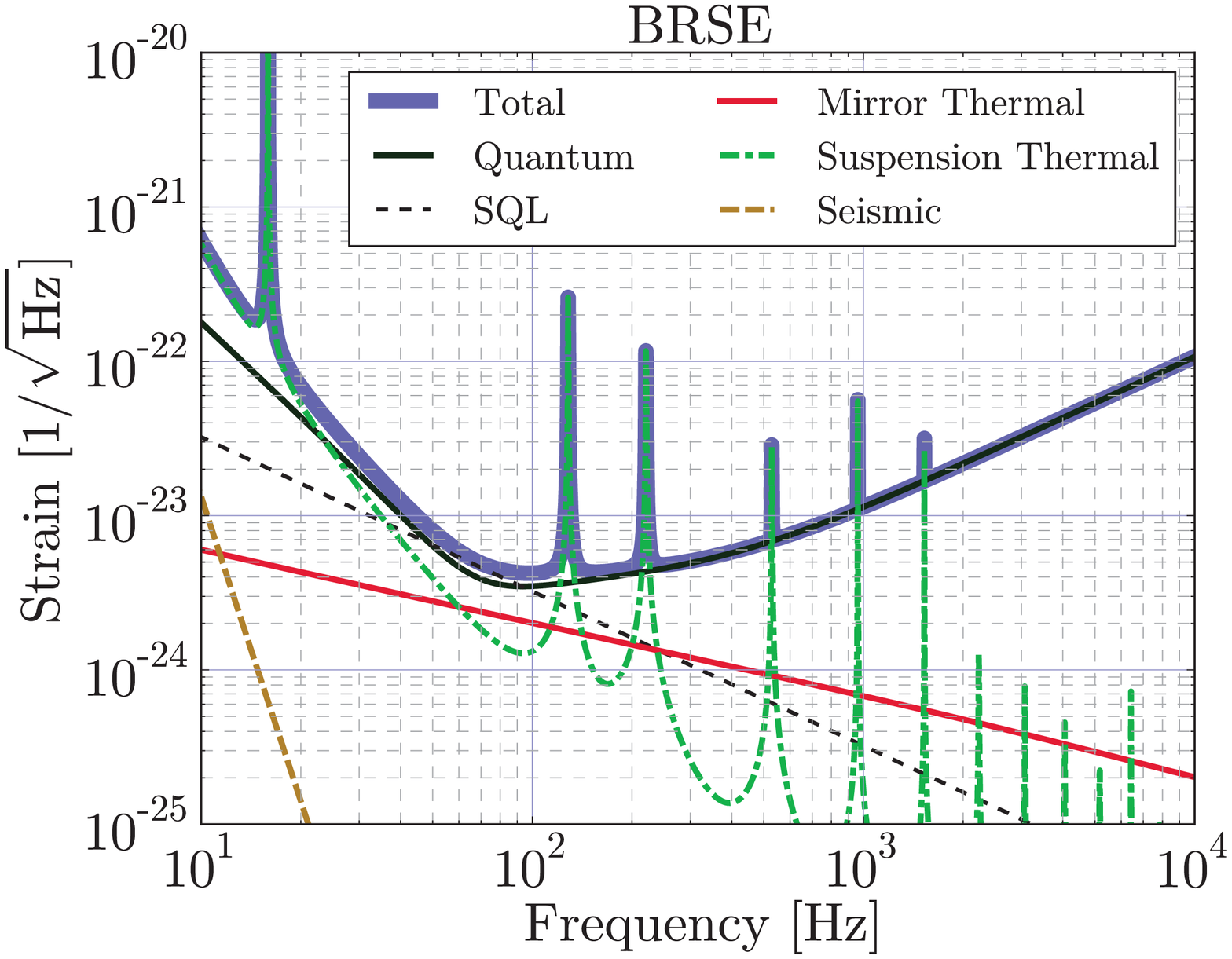}    
  \end{minipage}
\hspace{0.5cm}
  \begin{minipage}{8cm}
\includegraphics[width=8cm]{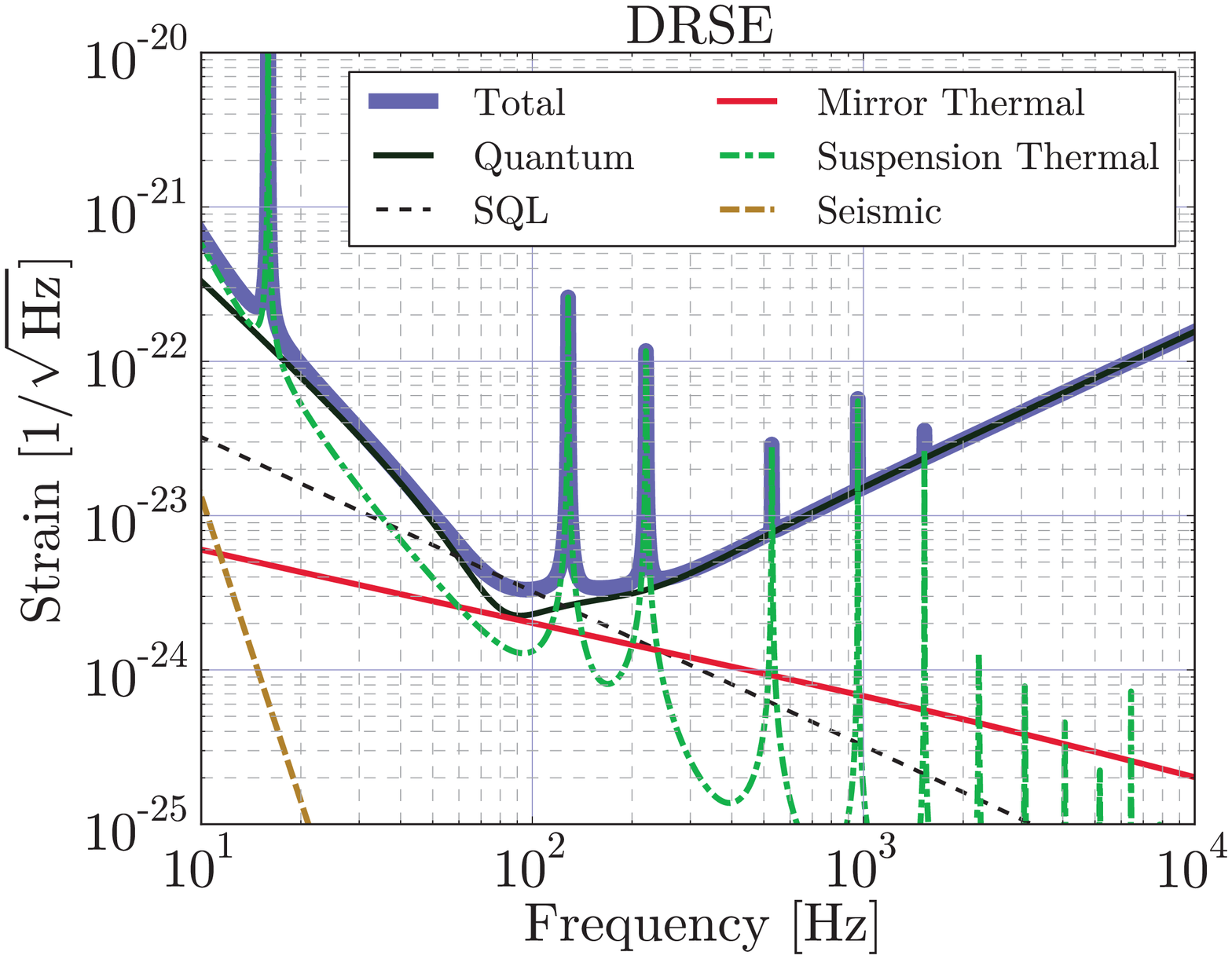}    
 \end{minipage}
\caption{Estimated noises of KAGRA. The total noise is the quadratic sum of all the noises.}
\label{fig:noise-budget}
\end{figure*}

In this section and the next section, we determine the target
sensitivities of KAGRA. As is explained in \ref{Section: QN Opt}, we
operate the detector with two different states of
SRC detuning. Therefore, there are two target sensitivity curves, corresponding
to the broad-band RSE (BRSE) configuration and the detuned RSE (DRSE)
configuration.

Figure\,\ref{fig:noise-budget} shows the estimated noises of the KAGRA
detector. The sensitivity is mostly limited by quantum noises.
Suspension thermal noise contributes to the total noise at low
frequencies (below 30\,Hz). In this section we first give an overview
of the non-quantum noises. Detailed discussion of those noises are given
in \cite{somiya_detector_2012}.

\subsection{Seismic noise}\label{sec:seismic}

Each sapphire test mass is suspended under a 2-story seismic
attenuation system (SAS) that combines a short and sturdy inverted
pendulum with a series of suspension stages with geometric anti-spring
(GAS) filters\,\cite{somiya_detector_2012}. In order to cool down the
mirrors, heat-links made of pure aluminum wires are attached to the
penultimate and upper stages of the suspension system. The auxiliary
mirrors are suspended by simpler suspension systems.

In general the seismic motion of the KAGRA site is very quiet: about
100 times lower than that of the TAMA site in
Tokyo\,\cite{lism_collaboration_ultrastable_2004}. However, the
seismic activity depends on season and weather. In order to estimate
the seismic noise of the interferometer mirrors, we used the simulated
transfer functions of the above mentioned suspension systems and the
measured ground vibration spectrum of a stormy day in the Kamioka
mine, which is a worst case scenario.

\subsection{Thermal noises}\label{sec:thermal}

\subsubsection{Heat extraction capacity}

In order to extract heat from the sapphire test masses, the mirrors
are suspended by sapphire wires, which have a high thermal
conductivity at low temperatures. Then the heat is transferred to the cold
heads of the cryocoolers by pure-aluminum wires connected to the upper
stages of the mirror suspension systems\,\cite{somiya_detector_2012}. 

In the current design, the expected heat absorption by an ITM from the
incident laser beam is about 1.2\,W\,\cite{somiya_detector_2012}. The
diameter of the sapphire wires is determined to be 1.6\,mm, so that 
this heat can be extracted without increasing the mirror temperature over 20\,K.

One notable change regarding the cooling system design from the one
explained in \cite{somiya_detector_2012} is that we now separate the
cooling paths for a test mass and radiation shields. In the previous
design, all of the four cryocoolers were connected to both the shields
and the mirror. In the current design, the test mass suspension is
connected to two cryocoolers and the radiation shields are connected
to the other two. This way, the heat absorbed by the radiation shields
coming from the large angle scattering of the mirror surface
does not affect the mirror temperature so much. This design
allows us to use more laser power, which is different from the value
used in \cite{somiya_detector_2012}.

\subsubsection{Mirror thermal noise}

The mirror thermal noise curve in Fig.~\ref{fig:noise-budget} is the
quadratic sum of substrate Brownian noise, coating Brownian noise, and
 substrate thermoelastic noise. Coating thermo-optic noise is
supposed to be very low at 20~K and is ignored here. The formulae and
the parameters used to calculate the thermal noises are given in
\cite{somiya_detector_2012}.

\subsubsection{Suspension thermal noise}

Calculation of the suspension thermal noise is performed using a
three-mass suspension system model consisting of a test mass, a
penultimate mass and a recoil mass suspended from the same penultimate
mass\,\cite{somiya_detector_2012}. The suspension materials for the
penultimate mass and the recoil mass are tungsten and copper
beryllium in this calculation.

The energy dissipation of a pendulum happens mainly at the top and
bottom ends of the suspension fibers. We used the average temperature
of the test and the penultimate masses as the effective temperature
for the calculation of the thermal noise from the horizontal
suspension modes. For the vertical modes, the effective temperature is
not trivial. We used the average temperature along the suspension
fiber for the calculation in this paper.

\section{Optimization of quantum noise shape}
\label{Section: QN Opt}

\subsection{Quantum non-demolition techniques}
\label{sec:quant-non-demol}

Quantum noises, i.e. shot noise and radiation pressure noise, are
mainly determined by input laser power, mirror reflectivities and
mirror masses. In addition to those parameters, we can modify the
quantum noise shape and possibly beat the standard quantum limit (SQL) by
using quantum non-demolition (QND) techniques. In KAGRA, we plan
to use two QND techniques. 

In order to extract the DARM signal, we use the DC readout
scheme\,\cite{ward_dc_2008} to avoid the shot noise penalty of the
conventional RF readout scheme and for many other practical
reasons\,\cite{somiya_frequency_2006}. In this scheme, a microscopic
offset in DARM is introduced during the operation to leak a weak
carrier field into the AS port. This carrier field serves as the local
oscillator for the gravitational wave sidebands (GWSBs) to generate
power variation proportional to the gravitational wave amplitude at
the AS port. In reality, there is also some carrier light leaking to
the AS port by the reflectivity difference of the two arm
cavities. The relative phase of the local oscillator to the GWSB,
called homodyne angle $\zeta$, is determined by the amplitude ratio of
these two carrier fields.  Therefore, by adjusting
the DARM offset, it is in principle possible to control the homodyne
angle. When an appropriate value of $\zeta(\ne 90^\circ)$ is chosen, a
cancellation of the shot noise and the radiation pressure noise
happens, beating the SQL. This QND technique is called back action
evasion (BAE).

The second QND technique to be employed in KAGRA is an optical spring
effect realized by detuning SRC\,\cite{buonanno_quantum_2001}. The
detuning imposes a rotation of the GWSB phase at the reflection by
the SRM. This induces differential radiation pressure force correlated to
the GWSB on the test masses, amplifying the GW signal at certain
frequencies. The parameter to characterize this QND scheme is the
detuning angle $\phi$ of the SRC, which is defined by $\phi\equiv 2\pi
d/\lambda$, where $d$ is the deviation of the SRC length from the
carrier resonance and $\lambda$ is the wavelength of the carrier
light.

\subsection{Optimization of the mirror reflectivities}
\label{sec:optim-mirr-refl}

The finesse $\mathcal{F}$ of the arm cavities and the reflectivity of
the SRM ($R_\mathrm{s}$) determine the quantum noise shape of an
interferometer together with the homodyne angle $\zeta$ and the
detuning angle $\phi$. These parameters are chosen by using the
detection range of binary neutron star inspiral events (Inspiral Range = IR) as a figure of
merit.

\begin{figure}[t]
	\begin{center}
		\includegraphics[width=8.5cm]{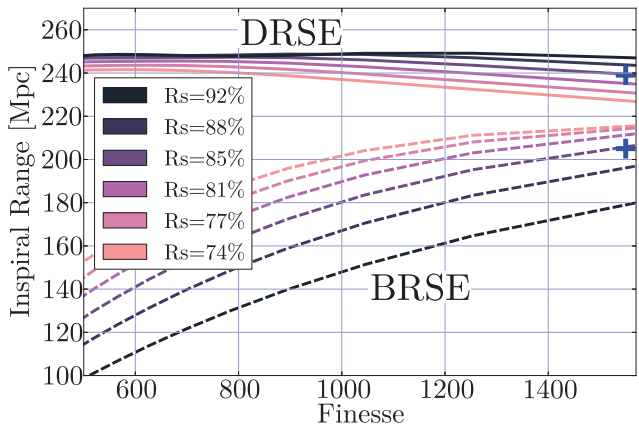}
	\caption{Inspiral range with different finesse and signal recycling mirror reflectivity.}
	\label{fig:IR}
	\end{center}
\end{figure}

Figure\,\ref{fig:IR} shows the inspiral range for $1.4_\odot -
1.4_\odot$ neutron star binaries as functions of $\mathcal{F}$ with
different values of $R_\mathrm{s}$. For each set of $\mathcal{F}$
and $R_\mathrm{s}$, $\zeta$ and $\phi$ are optimized to give the
largest IR. The input power is adjusted to make the heat absorption of
the test masses constant to keep the mirror temperature at 20\,K. We
also plotted the cases with $\phi=0$ to see the IRs for BRSE
configurations.  We assumed a round-trip loss of 100\,ppm for each arm
cavity for this calculation.

As is evident from the plot, DRSE configurations give generally better
inspiral ranges. However, a DRSE interferometer gives a narrower
detection bandwidth than the BRSE configuration of the same mirror
parameters. For example, in the case of
figure\,\ref{fig:noise-budget}, the sensitivity of BRSE is better than
DRSE above 500\,Hz, where signals from the merger phase of a neutron
star inspiral event are expected to
appear\,\cite{kiuchi_long-term_2009}\cite{rezzolla_accurate_2010}. This
means, for the first detection, a DRSE interferometer gives us a
better chance, while richer scientific information may be extracted
from a BRSE detector. Moreover, operation of a DRSE interferometer has
some technical concerns, such as unwanted error signal offsets in the
auxiliary DOFs by an imbalance of the RF sidebands used for signal
extraction. Therefore, fixing the interferometer configuration to DRSE
bears some risks. For those reasons, we decided to make our
interferometer to be operated in both configurations (variable
detuning). Variable detuning is realized by adding an offset into the
error signal for the control of the SRC length. Therefore, the amount
of possible detuning is limited to the linear range of the SRC error
signal.

From figure\,\ref{fig:IR}, we selected reflectivity parameters to have
good IRs for both BRSE and DRSE. These are indicated by + marks in the
plot. A higher finesse makes the interferometer susceptible to the
losses of the mirrors. This trend can be seen in the DRSE
curves. A smaller finesse decreases the BRSE sensitivity. Also, the
optimal detuning angles of smaller finesse cases are too large to be
realized by the offset detuning method explained above.  The value of
$R_\mathrm{s}$ is selected to strike a balance between BRSE and DRSE. The
PRM reflectivity is chosen to match the reflectivity of the arm
cavities.  Table\,\ref{tab: refl-params} summarizes the selected reflectivities
of the mirrors as well as the homodyne and the detuning angles of
the KAGRA interferometer.

\begin{table}
  \caption{Parameters of the KAGRA interferometer related to the quantum noises.\label{tab: refl-params}}

\begin{ruledtabular}
\begin{tabular}{ll|ll}
Arm cavity finesse&1530&ITM Reflectivity& 99.6\%\\
ETM Loss & $<$ 50\,ppm & PRM Reflectivity& 90\%\\
SRM Reflectivity & 85\%& Homodyne angle&$132^\circ$\\ 
Detuning angle&$3.5^\circ$&Input Laser Power&78\,W\\
BRSE IR&217Mpc&DRSE IR&237Mpc\\

\end{tabular}
\end{ruledtabular}
\end{table}


\section{Length sensing and control}
\label{Section: RFSB}

In order to operate the interferometer with the quantum noise limited
sensitivity discussed in the previous section, the interferometer
mirrors have to be kept at certain operation states. For example, the
arm cavity lengths have to be kept at an integral multiple of the
laser wavelength to resonate the light inside them. To achieve this,
the positions and the orientations of the mirrors have to be monitored
first. Then feedback control is used to keep them at the optimal
operating points throughout the operation of the interferometer.

In this section, we discuss how to extract necessary information to
control the mirrors. Although we have to control both positions and
the orientations of the mirrors, we only focus on the position (or
length) control in this section. The alignment control is discussed in
section \ref{Section: Alignment}. The full account of the length
signal extraction scheme is given in\,\cite{aso_length_2012} and this
section is a brief summary of the work.

\subsection{RF sidebands resonant conditions}
We mainly use a variant of the RF readout scheme to extract the
information of the most of the degrees of freedom (DOFs) to be
controlled. Only the DARM signal is obtained using the DC readout scheme
as discussed in the section \ref{Section: QN Opt}.

Our sensing scheme makes use of RF sidebands, generated by phase
modulations applied to the incident laser beam. Those RF sidebands
generate beat signals against the carrier or other RF sidebands at the
output ports of the interferometer. These beat signals contain
information on the motions of the mirrors, usually mixtures of various
DOFs. In order to extract the information of DOFs independently, the
RF sidebands have to see different parts of the interferometer, i.e. they
must resonate in different parts of the interferometer.

\begin{figure}
\includegraphics[width=8cm]{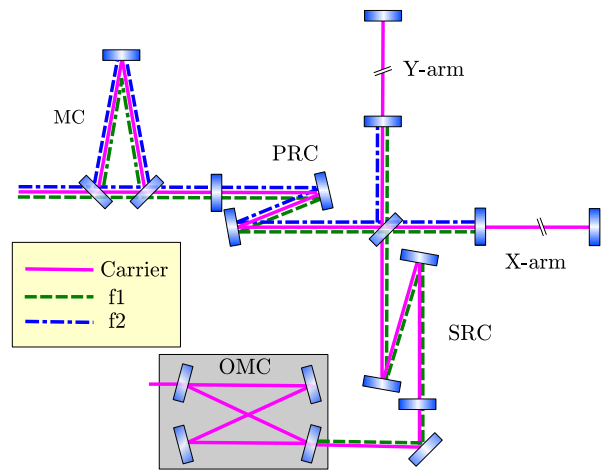}
\caption{Resonant conditions of the carrier and the RF sidebands. Each
  field is represented by lines of a distinct style. A field is
  resonant in the parts of the interferometer the corresponding lines are drawn.}
\label{Figure: Resonant Condition}
\end{figure}

For the interferometer control of KAGRA, we use two sets of RF
sidebands, called f1 and f2. The resonant conditions for those
sidebands and the carrier inside the interferometer are depicted in
figure \ref{Figure: Resonant Condition}. The carrier is resonant in
the two arm cavities and the PRC. The AS side of the BS is kept at a
dark fringe for the carrier. The f1 sidebands are resonant in the PRC
and the SRC, but not in the arm cavities. The f2 sidebands resonate
only in the PRC. In this way, we can expect those light fields to
carry different information of the mirror motions.  Since we apply the
modulations before the MC, the two RF sidebands also have to resonate
in the MC.

\subsection{Modulation frequencies and the macroscopic lengths}
\label{sec:rf-sb-freq-len}

The resonant conditions of the RF sidebands are determined by the
macroscopic lengths of the PRC and the SRC and the macroscopic
asymmetry of the MICH as well as the RF modulation frequencies.  There
are many combinations of those parameters which can realize the
resonant conditions explained above. However, we have to satisfy
several practical constraints in choosing them.

First of all, it is desirable to have short PRC and SRC from the view
point of construction cost, especially in the underground site of
KAGRA. However, the PRC and the SRC lengths have to be long enough to
house the Z shaped folding part without causing too much astigmatism
on the laser beams. In addition, we have to include 20\,m long cold
sections in the vacuum pipes between the ITMs and the BS to prevent
the room temperature thermal radiation from bombarding the cold test
masses.  The RF modulation frequencies are constrained to below 50MHz
from the available response speed of photo detectors with large
apertures. It is also desirable to be above 10MHz to avoid large
low-frequency laser noises.

We tested a large number of combinations of the length and the
frequency parameters to find ones which satisfy the resonant
conditions and the practical constraints at the same time. Out of
several survived candidates, we chose the parameters shown in Table
\ref{tab: len-freq-param}. We used the loop noise coupling estimates,
explained in the next section, to decide the best parameter
sets\,\cite{aso_length_2012}.

\begin{table}
\caption{Length and frequency parameters. The values shown here are after the adjustment of section\,\ref{sec:fine-tuning-rf}. \label{tab: len-freq-param}}

\begin{ruledtabular}
\begin{tabular}{ll|ll}
Arm cavity length&3000\,m&f1 frequency& 16.881MHz\\
PRC Length & 66.591m & f2 frequency& 45.016MHz\\
SRC Length & 66.591m& MC Length&26.639\,m\\ 
Michelson asymmetry&3.30\,m&&\\

\end{tabular}
\end{ruledtabular}
\end{table}

\subsection{Fine tuning of the RF sideband frequencies}
\label{sec:fine-tuning-rf}

The RF sidebands f1 and f2 are almost anti-resonant to the arm
cavities but not perfectly so. A consequence of this is that they get
small but finite phase shifts when reflected by the arm
cavities. Those two sidebands have to resonate in the PRC at the same
time. However, if the phase shifts they get from the arm cavities are
arbitrary, the resonant conditions for them is different, thus we can
not resonate both of them at the same time.  For this reason, a fine
tuning of the RF sideband frequencies is necessary.

An effective cavity length change caused by a phase shift $\phi$ for a
modulation sideband with a modulation frequency $\omega_\mathrm{m}$ is
$\Delta L = \phi c/\omega_\mathrm{m}$. Therefore, if the phase shifts for
the f1 and f2 sidebands are proportional to their frequencies, the effective
length change is the same for the two SBs. Then we can just
pre-shorten the PRC length by this amount to fulfill the resonant
conditions for both of the sidebands at the same time.

\begin{figure}[tbp]
\includegraphics[width=8cm]{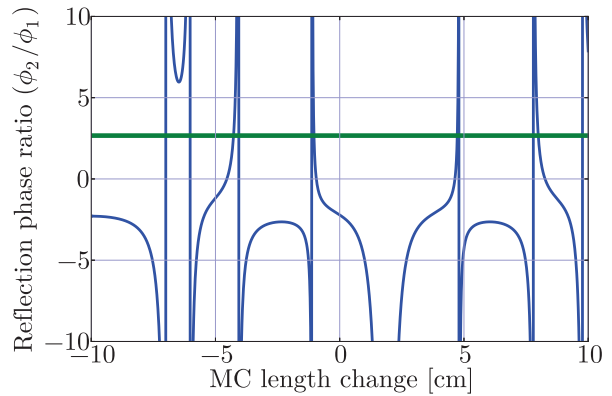}
\caption{Ratio of the RF sideband reflection phases by the arm cavities. We want
  to set it to 8/3, which is indicated by the green line.}
\label{RFSB Refl phase ratio}

\end{figure}

In order to adjust the reflection phases for f1 and f2, we need to
change their frequencies relative to the carrier resonance. However,
we have to keep the ratio of f1 and f2 frequencies to be 3:8 to
fulfill the resonant conditions of figure \ref{Figure: Resonant
  Condition}.  This is automatically satisfied by requiring the two
sidebands to transmit the MC, i.e. the f1 frequency is 3 times the
free spectral range (FSR) of the MC and f2 is 8 times the MC
FSR. Therefore, we will slightly change the MC length from its nominal
value to find the optimal RF sideband frequencies which give the
desired arm cavity reflection phases. The precise amount of phase
shifts induced on nearly-anti-resonant fields by a cavity depends on
its finesse.  Therefore, the RF sideband frequencies must be adjusted
according to the measured value of the real arm cavity finesse. In
this paper, we assume 100\,ppm of loss in the arm, resulting in the
finesse of 1530.  Figure\,\ref{RFSB Refl phase ratio} shows the ratio
of the reflection phases ($\phi_2/\phi_1$) as a function of the MC
length. The desired value of 8/3 is indicated by a green horizontal
line.  By finding an intersection of the blue curve with the green
line, tentative numbers for the RFSB frequencies are determined to
be f1=16.881\,MHz and f2=45.016\,MHz. Corresponding changes of the PRC
and the SRC lengths are 5.7\,mm and 11.4\,mm respectively.

\subsection{Sensing matrix}
\label{sec:sensing-matrix}

Once a parameter set is chosen for the macroscopic length and the
modulation frequencies, we can calculate the response of the
interferometer, that is, how much beat signals at each detection
port is changed in response to the motion of the mirrors. We used an
interferometer simulation tool, called
Optickle\,\cite{evans_optickle_2007}, for this calculation. By solving
self-consistent equations of optical fields inside the
interferometer, Optickle computes the strength of the beat signals at
each detection port. Optickle also takes into account the radiation
pressure effect to correctly compute the response of a high power
interferometer.

There are three detection ports and two primary beat frequencies (f1
and f2). For each beat frequency, there is a choice of two orthogonal
demodulation phases. Therefore, there are $3\times 2\times 2=12$
candidate signals to be used as error signals. Out of those signals,
we chose ones with good signal strength and separation between DOFs.
The signal sensing matrice of the KAGRA interferometer with the
selected parameters and signal ports are shown in Table\,\ref{Table:
  LSCSensMatBRSE} and \ref{Table: LSCSensMatDRSE}.  These
matrice are frequency dependent. The values shown in the tables are
calculated at 100\,Hz.

\begin{table*}
  \caption{Length sensing matrix for BRSE. {\bf AS\_DC}: DC readout signal at the AS port. {\bf REFL\_f1I}: REFL signal demodulated at the f1 frequency in in-phase. {\bf REFL\_f1Q}: REFL signal demodulated at the f1 frequency in quadrature-phase. {\bf POP\_f1I} and {\bf POP\_f2I}: POP signal demodulated in in-phase at the f1 and the f2 frequencies respectively. The values are the transfer coefficients at 100\,Hz from the motion of the mirrors to the signal ports, with each row normalized by the diagonal element. \label{Table: LSCSensMatBRSE}}
  \begin{tabular}{|c@{\ \vrule width 0.8pt\ }c|c|c|c|c|}
\hline
&   {\bf DARM} &{\bf CARM} &{\bf MICH} &{\bf PRCL}&{\bf
   SRCL}\\\noalign{\hrule height 0.8pt}
   & & & & & \\[-10pt] 
{\bf AS\_DC}&1&$4.2\times10^{-5}$&$1.0\times10^{-3}$&$4.8\times10^{-6}$&$4.7\times10^{-6}$\\\hline
{\bf
   REFL\_f1I}&$5.4\times10^{-3}$&1&$4.3\times10^{-5}$&$6.5\times10^{-3}$&$4.3\times10^{-3}$\\\hline
{\bf REFL\_f1Q}&$5.0\times10^{-3}$&$1.3\times10^{-2}$&1&1.02&0.67\\\hline
{\bf
   POP\_f2I}&$2.3\times10^{-2}$&4.3&$1.0\times10^{-2}$&1&$2.5\times10^{-4}$\\\hline
{\bf POP\_f1I}&$8.7\times10^{-2}$&16.2&$3.1\times10^{-2}$&2.1&1\\\hline
  \end{tabular}
\end{table*}

\begin{table*}
  \caption{Length sensing matrix for DRSE. {\bf REFL\_f2I}: REFL signal demodulated at the f2 frequency in in-phase. \label{Table: LSCSensMatDRSE}}
  \begin{tabular}{|c@{\ \vrule width 0.8pt\ }c|c|c|c|c|}
\hline
&   {\bf DARM} &{\bf CARM} &{\bf MICH} &{\bf PRCL}&{\bf
   SRCL}\\\noalign{\hrule height 0.8pt}
   & & & & & \\[-10pt] 
{\bf AS\_DC}&1&$4.1\times10^{-5}$&$1.0\times10^{-3}$&$4.5\times10^{-6}$&$7.6\times10^{-6}$\\\hline
{\bf
   REFL\_f2I}&$1.2\times10^{-2}$&1&$1.3\times10^{-4}$&$1.2\times10^{-2}$&$1.4\times10^{-3}$\\\hline
{\bf REFL\_f1Q}&$2.8\times10^{-2}$&$9.9\times10^{-3}$&1&0.39&0.18\\\hline
{\bf
   POP\_f2I}&$2.7\times10^{-2}$&4.3&$1.0\times10^{-2}$&1&$8.5\times10^{-5}$\\\hline
{\bf POP\_f1I}&$1.7\times10^{-1}$&35&$3.1\times10^{-2}$&2.0&1\\\hline
  \end{tabular}
\end{table*}

\subsection{Loop noise coupling}
\label{sec:loop-noise-coupling}

The sensing matrice shown in Table\,\ref{Table: LSCSensMatBRSE} and
\ref{Table: LSCSensMatDRSE} are clearly not diagonal. Therefore,
feedback control using these signals causes some cross-couplings
between the DOFs. Since each error signal has its own noises, the
cross-couplings could inject excess noises from the auxiliary DOFs
into the DARM signal. This mechanism is called loop noise
coupling\,\cite{somiya_shot-noise-limited_2010}. Along with the
interferometer response to the mirror motions, Optickle can also
compute the quantum noise at each signal port. Using this information, we
can compute the amount of loop noise couplings.

For the calculation of the loop noise, we have to assume the shape of
open-loop transfer functions of the servo loops. We assumed a simple
$1/f^2$ shaped transfer function with $1/f$ response around the unity
gain frequency (UGF) for stability. The UGFs are set to 200\,Hz for
DARM, 10\,kHz for CARM and 50\,Hz for all the other DOFs.

\begin{figure*}
  \begin{minipage}{8cm}
\includegraphics[width=8cm]{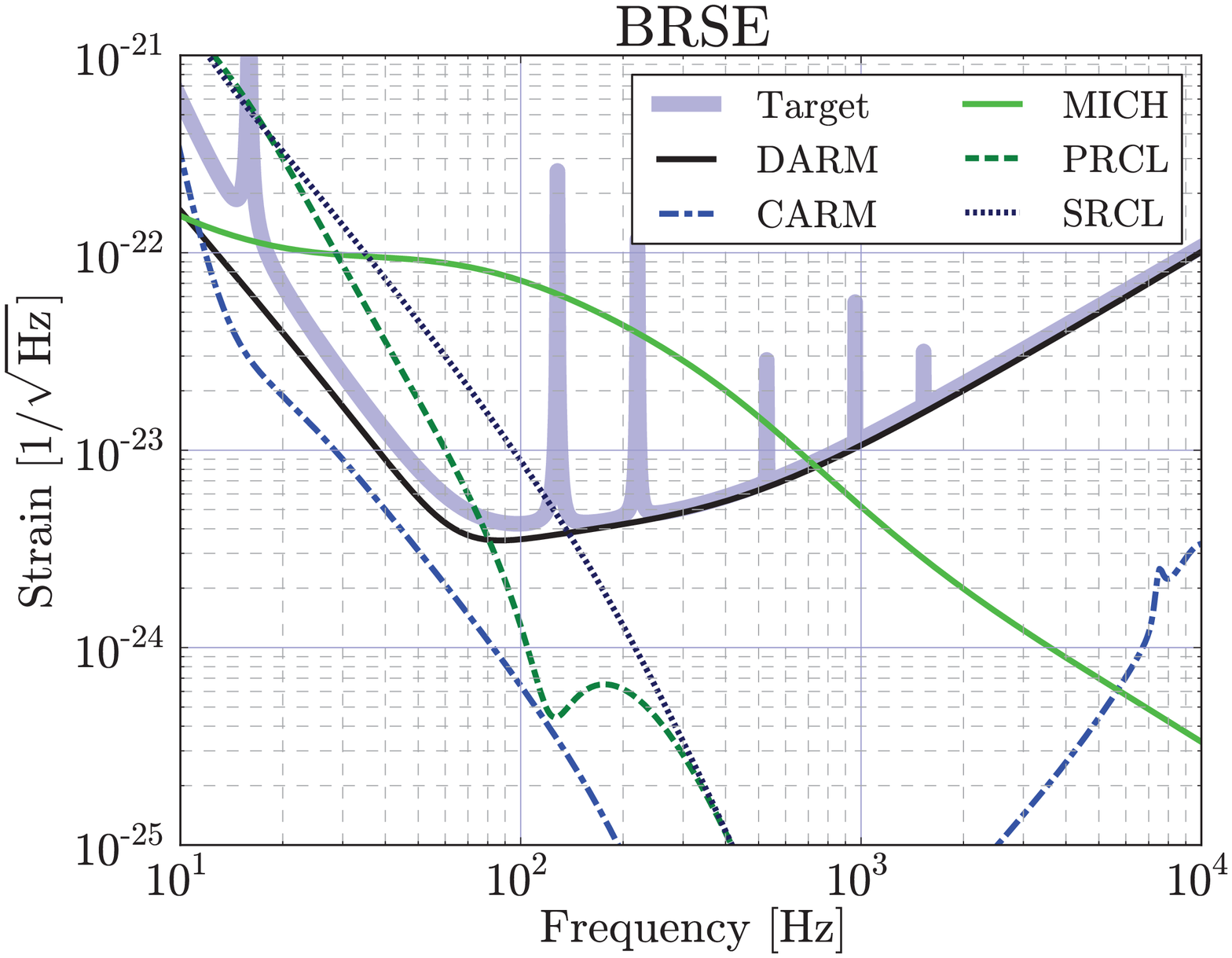}    
  \end{minipage}
\hspace{0.5cm}
  \begin{minipage}{8cm}
\includegraphics[width=8cm]{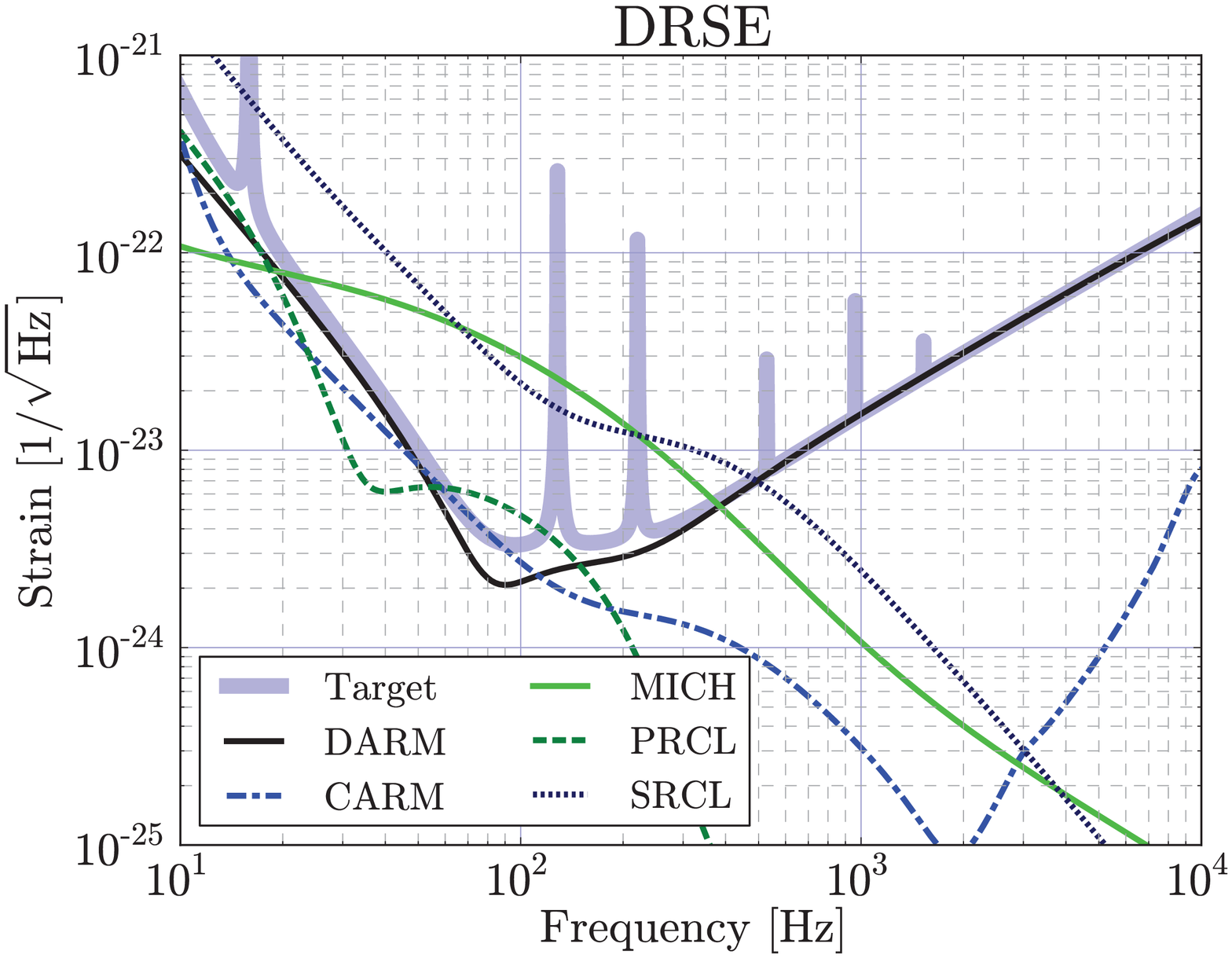}    
 \end{minipage}
\caption{Loop noise couplings}
\label{Figure: LSC Loop Noise}
\end{figure*}

The calculated loop noise couplings to the DARM signal are shown in
figure \ref{Figure: LSC Loop Noise}. Obviously, the injected noises
from the auxiliary DOFs are too high and compromising the target
sensitivity. However, we can mitigate this problem by using a
technique called feed-forward\,\cite{aso_length_2012}, which was
widely used in the first generation interferometric detectors. A
loop noise is first injected into an auxiliary mirror, such as the PRM, by
a feedback force. Then the noise-induced motion of this mirror is
coupled to the DARM signal by the off-diagonal elements of the sensing
matrix. This means that by knowing how much force is applied to the
auxiliary mirrors, and by experimentally measuring the
off-diagonal elements of the sensing matrix, we can precisely estimate
the noise signal injected into DARM. Then, we can subtract this noise
either by signal processing or by feed-forwarding the estimated noise
to the DARM actuators with opposite sign.

\begin{figure*}
  \begin{minipage}{8cm}
\includegraphics[width=8cm]{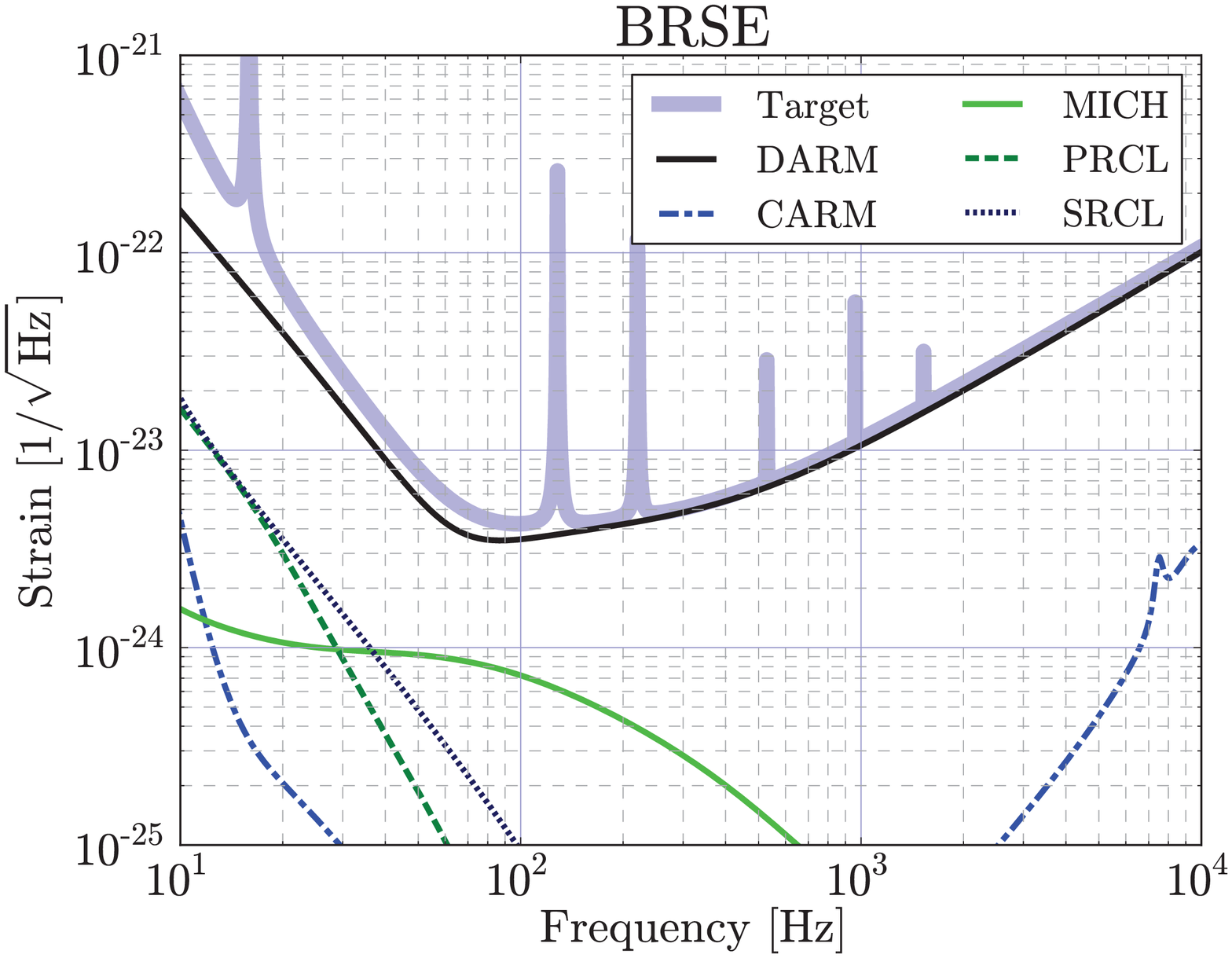}    
  \end{minipage}
\hspace{0.5cm}
  \begin{minipage}{8cm}
\includegraphics[width=8cm]{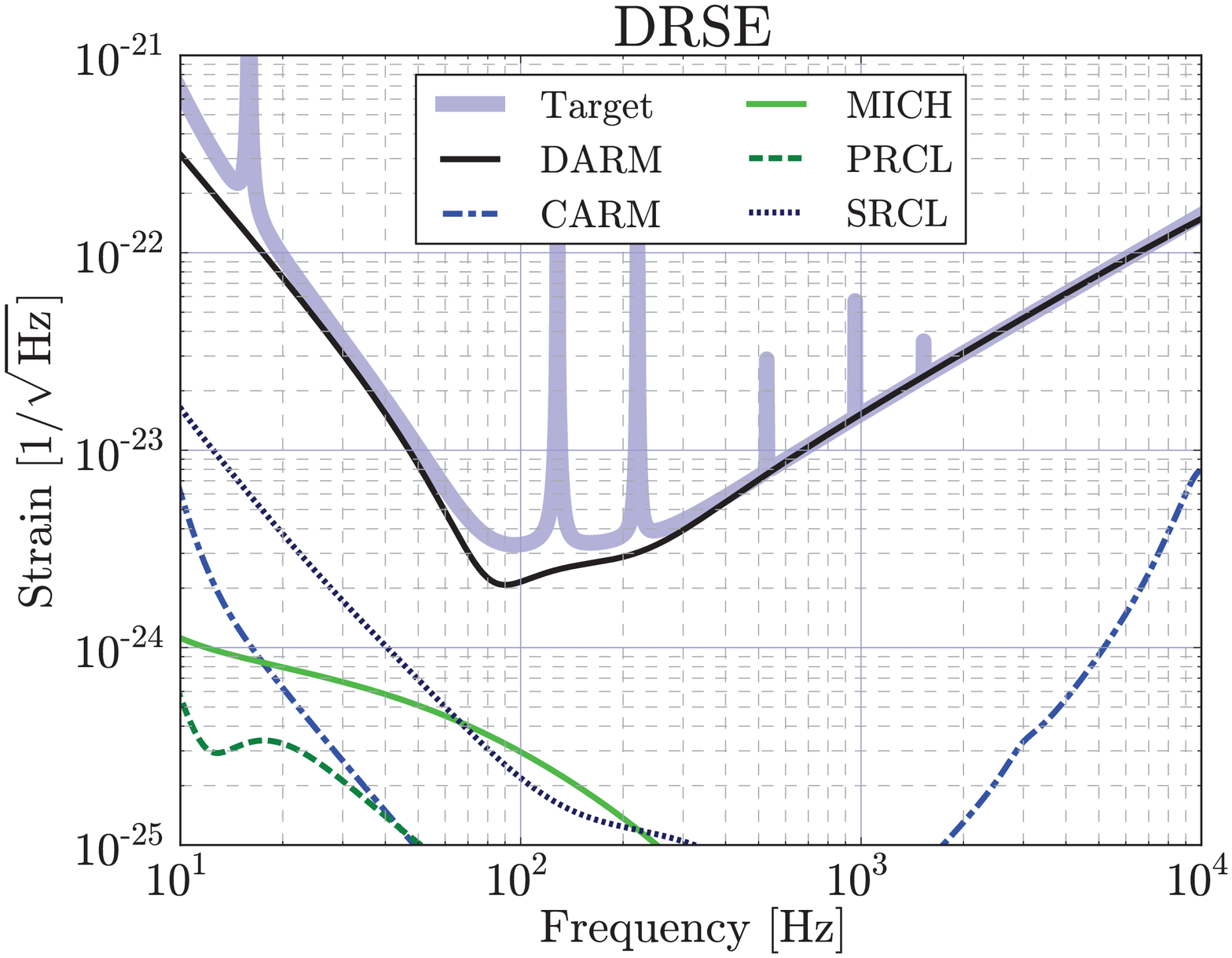}    
 \end{minipage}
\caption{Loop noise couplings with feed-forward.}
\label{Figure: LSC Loop Noise with FF}
\end{figure*}

Figure\,\ref{Figure: LSC Loop Noise with FF} shows loop noise
couplings after the feed-forward scheme is applied. We assumed that the
noise cancellation is performed with 1\% accuracy. This result assures
that the signal sensing scheme and the parameters we selected yield
sufficiently low-noise signals for the length control of the
interferometer. In the design process, we repeatedly computed the loop
noise plots like figure \ref{Figure: LSC Loop Noise with FF} with
various interferometer parameters to choose the best set of the
parameters.

\section{Spatial optical modes}
\label{Section: Spatial modes}

Up to this point, we analyzed the interferometer with the scalar field
approximation, disregarding the spatial mode shape of the laser
beams. However, in a real interferometer, we have to mode match the
various parts of the interferometer to resonate only the necessary
optical mode, i.e. TEM00 mode. In this section, we consider
the spatial mode design of the interferometer with the goal of  determining the radii
of curvature of the interferometer mirrors.

\subsection{Arm cavities}
\label{sec:arm-cavities-1}

\subsubsection{g-factor}
\label{sec:g-factor}

The spatial mode characteristics of a Fabry-Perot cavity is determined
by the g-factors, which are defined as follows:
\begin{equation}
  \label{eq:1}
  g = g_1\cdot g_2, \quad g_1\equiv 1-\frac{L}{R_1},\quad g_2\equiv 1-\frac{L}{R_2},
\end{equation}
where $L$ is the length of the cavity and $R_1$ and $R_2$ are the ROCs
of the ITM and the ETM. The g-factors determine the beam spot sizes on
the mirrors and the degree of degeneracy of the higher-order spatial modes
in the cavity.

First, we consider the beam spot sizes on the mirrors, because they
directly affect the noise of the interferometer through the thermal
noise coupling. We want to make the beam sizes as large as possible to
reduce mirror thermal noises.  If we assume $R_1 = R_2$, the beam
spot size $w$ is the same on both the mirrors, and it can be written
as a function of the common g-factor $g_0 \equiv g_1 = g_2$,

\begin{equation}
  \label{eq:2}
  w = \sqrt{\frac{\lambda L}{\pi}\sqrt{\frac{1}{(1+g_0)(1-g_0)}}}.
\end{equation}
Since this is an even function of $g_0$, there are two possible values of $g_0$
giving the same spot size.

The high optical power circulating inside KAGRA's arm cavities
generates strong angular optical spring
effects\,\cite{sidles_optical_2006}. There are always two angular
spring modes: one is a positive-spring and the other is a negative-spring
mode. The negative-spring mode causes angular instability of the
mirrors if it is stronger than the mechanical restoring force of the
mirror suspension. It is known that for the same beam size, the
negative-spring constant is made smaller by choosing a negative
g-factor ($g_0$). Therefore, we prefer a negative value of
$g_0$.

For our 22\,cm diameter mirrors, the maximum possible beam size is
4.0\,cm requiring the diffraction loss per reflection to be less than
1\,ppm. The negative $g_0$ to realize this spot size is -0.772,
corresponding to the mirror ROC of 1692\,m. However, because of the
time and cost constraints, we had to choose an ROC which can be
polished using one of the stock reference spheres of the polishing
company. For this reason, we had to change the ROC to 1900\,m. This
reduces the beam spot size to 3.5\,cm. The thermal noise increase by this
change degrades the IR from 221\,Mpc to 217\,Mpc for BRSE and from
243\,Mpc to 237\,Mpc for DRSE.

\subsubsection{Carrier higher order mode resonances}
\label{sec:higher-order-mode}

Ideally, the arm cavities should resonate only the TEM00 mode during
the operation. However, optical higher order modes (HOMs) are not
completely anti-resonant to the arm cavity in general.  Therefore, if
there is mis-alignment or mode mis-matching, HOMs could resonate in
the arm cavities, potentially increasing the shot noise. If the
selected arm g-factor is a particularly bad one, this HOM coupling
could be large.  In this section, we confirm that our g-factor does
not allow excessively large resonances of HOMs.

\begin{figure}[tbp]
\includegraphics[width=8cm]{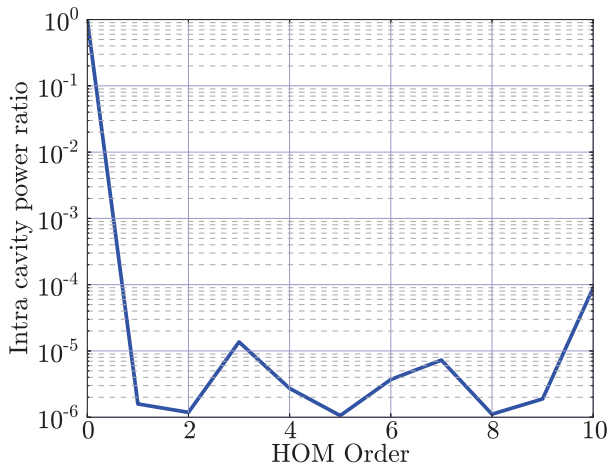}
\caption{HOM power in the arm cavity relative to the TEM00 power. The
  mode number is $n+m$ for TEMnm modes.}
\label{HOM Power in AC}
\end{figure}

Figure\,\ref{HOM Power in AC} shows the HOM power ratio to the TEM00
power in an arm cavity. This is the ratio of the intra-cavity optical
power, if TEM00 and HOM modes are injected to the arm cavity with
the same power. When calculating an HOM power, we took into account
the fact that for HOMs, the diffraction loss is higher than for
TEM00. This is because HOMs are spatially spread more widely. The
diffraction losses were calculated with an FFT optical simulation
tool SIS\,\cite{yamamoto_sis_2007}.

\begin{figure}[tbp]
\includegraphics[width=8cm]{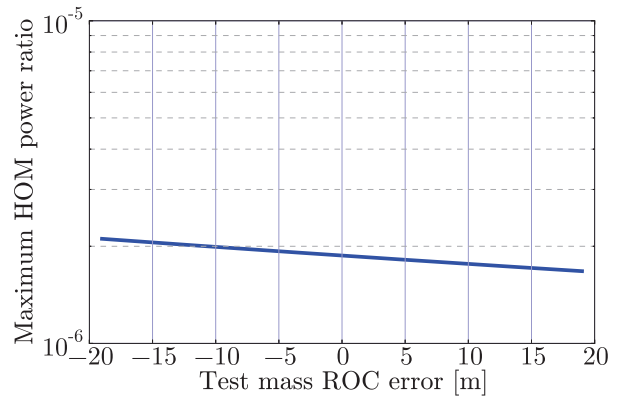}
\caption{The maximum HOM power ratio in the arm cavity as a function of the test mass ROC
  error. The ROC is swept by $\pm$1\% around
  the nominal value.}
\label{Max HOM Power}
\end{figure}

Figure\,\ref{HOM Power in AC} assumed that the g-factor of the cavity
is exactly as designed. In reality, there is always some error in the
ROCs of real mirrors. We set the error tolerance to be $\pm$0.5\%
mainly from the technical feasibility of mirror polishing.
Figure\,\ref{Max HOM Power} shows the maximum HOM power ratio (the
value of the highest peak in Figure\,\ref{HOM Power in AC} except for
the mode number = 0) as a function of ROC error. There is no
significant change in the HOM power ratio throughout the error
range. This means that our g-factor is robust against mirror
fabrication errors.

\subsubsection{RF sideband higher order resonances}
\label{sec:rf-sideband-higher}

Although the RF sideband frequencies are chosen to be not resonant to
the arm cavities for TEM00 mode, their HOMs may accidentally hit a
resonance of the arm cavities. This can cause an unwanted coupling of
arm cavity alignment fluctuations to the error signals of the
auxiliary DOFs.

\begin{figure}[tbp]
\includegraphics[width=8.5cm]{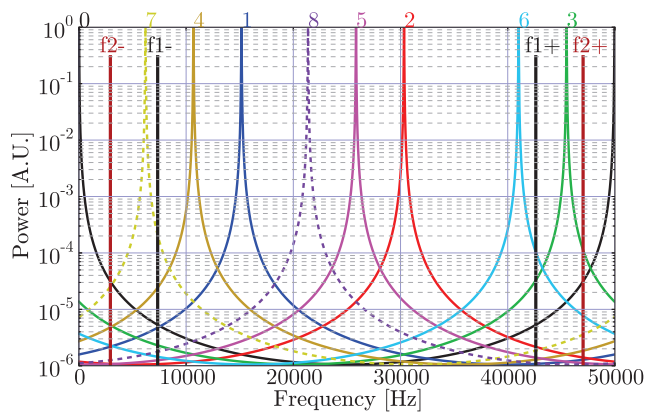}
\caption{Positions of the RF SBs and the HOMs in an FSR of the arm
  cavities. The colorful sharp peaks represent the resonant curves of
  the HOMs. The mode numbers are printed at the top of each resonance. The
  vertical lines are the positions of the RF SBs. The ``+'' and ``-'' signs indicate upper and lower sidebands, respectively.}
\label{RFSB HOM FSR}
\end{figure}

Figure\,\ref{RFSB HOM FSR} shows the positions of the RF
sidebands and their HOMs in the FSR of the arm cavity. In the figure,
both the HOM resonant curves (Lorentzian-shaped curves with mode
numbers) and the frequencies of the RF sidebands (vertical lines) are
shown. We can see that there is no significant overlap between the RF
sidebands and the HOM resonances.

In reality, the exact frequencies of the RF sidebands change according
to the fine tuning, explained in section
\ref{sec:fine-tuning-rf}. Therefore, accidental coincidence of an RF
sideband and an HOM resonance could still happen. In this case, we can
try to use a different crossing point of figure\,\ref{RFSB Refl phase
  ratio} to move the RF sideband frequencies away from problematic HOM resonances
and avoid unfortunate overlaps.

\subsection{Recycling cavities}
\label{sec:recycling-cavities}

\subsubsection{Gouy phase shifts}
\label{sec:gouy-phase-shifts}

Now we turn our attention to the recycling cavities. The spatial mode
characteristic of the recycling cavities can be determined by the Gouy
phase changes of the light along the optical paths of the
cavities. Because the arm cavities are very long, if we use a straight
recycling cavity and inherit the spatial mode of the arm cavities
without modification, the one-way Gouy phase change inside the
recycling cavity is very small (less than 1 degree). This makes the
cavity highly degenerated for HOMs. Therefore, small alignment
fluctuations or thermal lensing can cause the excitation of HOMs
inside the recycling cavities. This is especially a problem for the RF
sidebands, which only resonate in the recycling cavities and do not
receive a mode healing effect from the stable arm cavities. The consequence is
poor spatial mode overlap between the RF sidebands and the carrier,
resulting in increased shot noise for the error signals of the
auxiliary DOFs. This was one of the most serious problems the first
generation interferometers struggled against.

In order to avoid the degenerated recycling cavity problem, we want to
increase the Gouy phase shifts in the recycling cavities. For this
purpose, we fold the cavities with two additional mirrors as shown in
figure\,\ref{Figure: IFO Schematic}. Before going into the details of
the folding scheme, first we discuss the desired values of one-way
Gouy phase changes in the PRC ($\eta_\mathrm{p}$) and the SRC
($\eta_\mathrm{s}$).

\subsubsection{Higher order mode power in the PRC}
\label{sec:degeneracy-prc}

\begin{figure}[tbp]
\includegraphics[width=9cm]{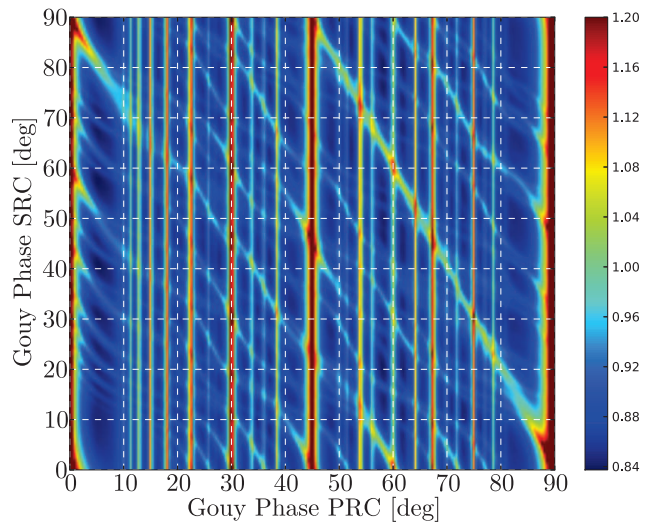}
\caption{Higher order mode resonance map of the PRC.}
\label{fig: HOM Resonance map PRC}
\end{figure}

Figure\,\ref{fig: HOM Resonance map PRC} shows a two-dimensional map
of the HOM degeneracy in the PRC. Each point in the map represent a
combination of ($\eta_\mathrm{p}$, $\eta_\mathrm{s}$). The color-coded
value at each point represents the severity of the HOM degeneracy and
it is computed as follows: Assuming the input laser power of 1W at the
carrier frequency injected from the back of the PRM, we compute the light power
circulating in the PRC by solving the static field equations of the
interferometer.  We repeat this calculation by changing the mode of
the input beam from TEM00 mode to HOMs of up to the 15th order. We
then take the sum of the computed intra-PRC power of the HOMs and
normalize it with the power of the TEM00 mode. If some of the HOMs are
close to the resonance in the PRC, this value (called
$\zeta_\mathrm{c}$) becomes large. This process is also repeated for
the f1 and the f2 sidebands, yielding the ratios $\zeta_\mathrm{f1}$
and $\zeta_\mathrm{f2}$. The color-coded value in the map of
figure\,\ref{fig: HOM Resonance map PRC} is the sum of
$\zeta_\mathrm{c}$, $\zeta_\mathrm{f1}$ and $\zeta_\mathrm{f2}$.


There are several dark areas in the map of figure\,\ref{fig: HOM
  Resonance map PRC}. We want to select $\eta_\mathrm{p}$ and
$\eta_\mathrm{s}$ centered in one of those large dark areas.  We did
not take the lower left area centered around $(\eta_\mathrm{p},
\eta_\mathrm{s})=(5.5^\circ, 8^\circ)$, because this region gives too
much degeneracy in the SRC, as explained in the next section.  In
order to select a preferred parameter set from the other candidate
areas, we computed the wave-front sensing signals (section
\ref{Section: Alignment}) and their diagonalized shot noises (equation
(\ref{eq:WFS shot noise})) repeatedly with all the candidate
parameters. As a consequence, we arrived at $\eta_\mathrm{p} =
16.5^\circ$ and $\eta_\mathrm{s} = 17.5^\circ$ as the parameter set to
give the lowest shot noises. The dark region around this parameter set
is not so large compared with other dark areas. However, the area is
sufficiently large considering the error tolerance of the ROCs of the
folding mirrors, discussed in section\,\ref{sec:roc-error-recovery}.

\subsubsection{Degeneracy of the SRC}
\label{sec:degeneracy-src}

In the previous section, we basically computed the HOM resonances in
the coupled PRC-SRC for the fields injected from the PRM. However,
there is one important process, which requires a separate treatment,
involving the degeneracy of the SRC. When there is a gravitational
wave passing through the detector, gravitational wave sidebands
(GWSBs) are excited in the arm cavities with opposite phases. These
GWSBs come out of the anti-symmetric side of the BS and reflected by
the SRM. If there is some figure error or defects on the surface of
the SRC mirrors, the GWSBs, which is in the TEM00 mode defined by the
arm cavities, can be scattered into HOMs inside the SRC. Although the
amount of the scattering by high quality mirrors is expected to be
very small, if one of the HOMs is resonant in the SRC, the scattering
loss is significantly enhanced\,\cite{pan_optimal_2006}, causing the
reduction of the net GW signal. This process can be investigated by
injecting a laser beam from the back of the SRM and checking the HOM
resonances.

We used the same field equations used in the previous section to
compute the SRC degeneracy, but injected the input beam from the back
of the SRM this time. The SRC length is controlled to be resonant to the
carrier by itself. However, when the arm cavities are locked, the
carrier gets an extra sign-flip at the back of the ITMs. Therefore,
they are not resonant in the SRC when the field is injected from the
SRM side. On the other hand, the HOMs of the carrier are not resonant
in the arm cavities, receiving no sign-flip from them. Therefore, a
degenerate SRC can resonate HOMs. Because of this resonant
conditions, it does not make sense to normalize the HOM powers with the
power of the TEM00 mode for this study. Instead, we first compute the
power of the n-th HOM, $P_\mathrm{d}(n)$, in the SRC when it is
completely degenerated, i.e. $\eta_\mathrm{s}=0$. Then we calculate
the same HOM power values, $P_\mathrm{s}(n)$, with a finite value of
$\eta_\mathrm{s}$. For each HOM, we take the ratio of the intra-SRC
power $\xi_n = P_\mathrm{s}(n)/P_\mathrm{d}(n)$. This ratio represents
how much the resonantly enhanced scattering problem is relieved by
adding a finite Gouy-phase shift to the SRC.

\begin{figure}[tbp]
\includegraphics[width=8cm]{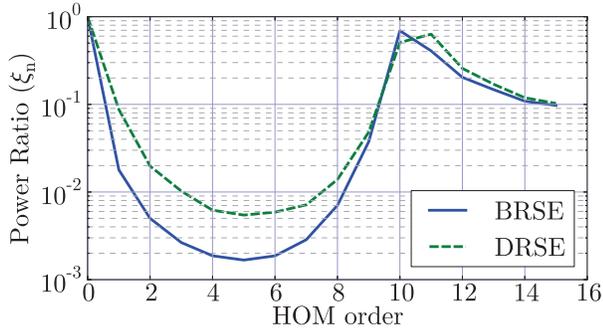}
\caption{Higher order mode resonance scan of the SRC for $(\eta_\mathrm{p},
\eta_\mathrm{s})=(16.5^\circ, 17.5^\circ)$.}
\label{fig: HOM Resonance scan SRC}
\end{figure}
 
Figure\,\ref{fig: HOM Resonance scan SRC} shows the computed $\xi_n$
as a function of the HOM order $n$ for $(\eta_\mathrm{p},
\eta_\mathrm{s})=(16.5^\circ, 17.5^\circ)$. For all the HOMs computed,
$\xi_n$ is smaller than 1, meaning the HOM resonance is reduced from
the completely degenerated case. At orders 10 and 11, the reduction is
not so large. However, we should note that because of the diffraction
loss, the finesse of the SRC is reduced by a factor of 3 for those
HOMs.

\begin{figure}[tbp]
\includegraphics[width=8cm]{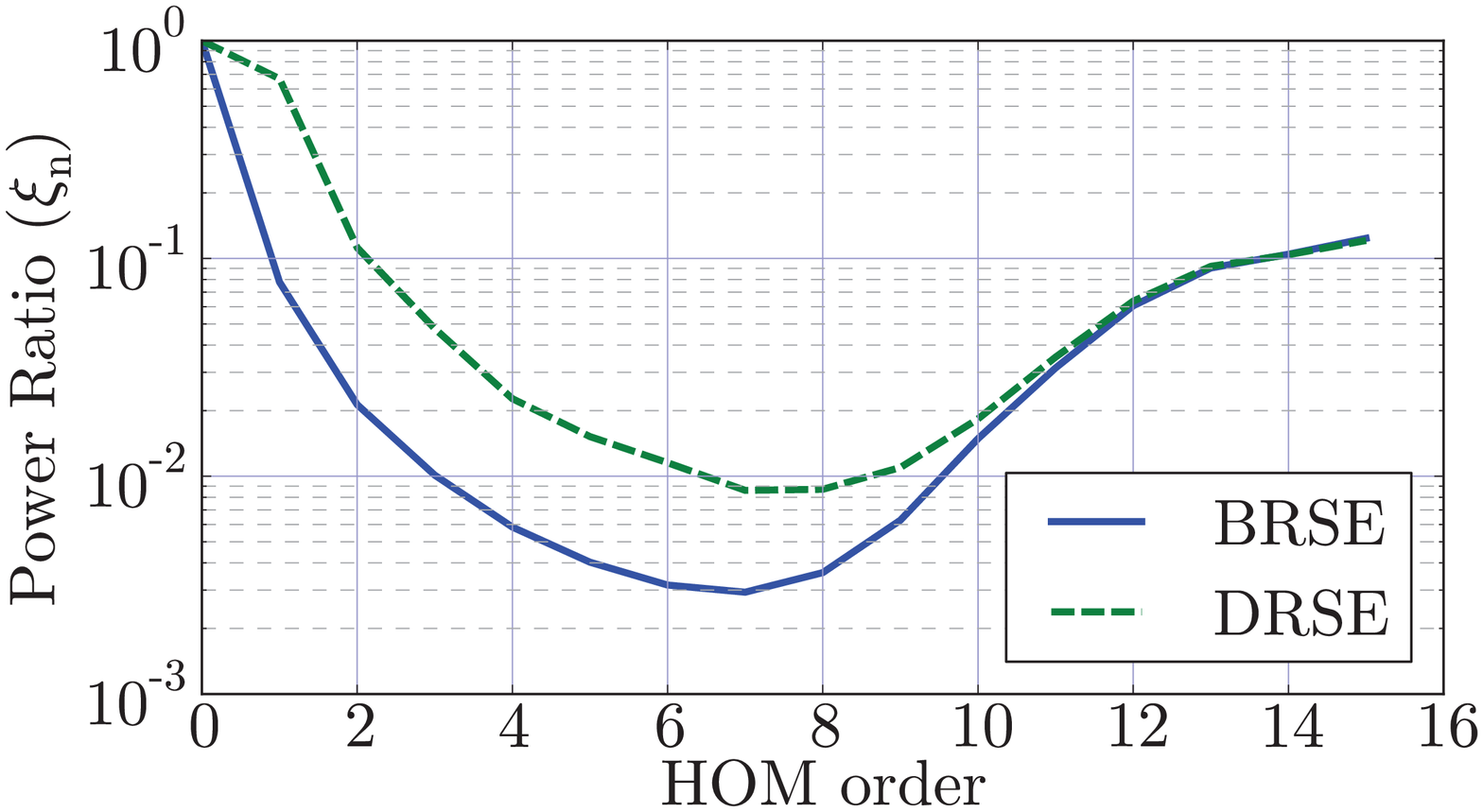}
\caption{Higher order mode resonance scan of the SRC for $(\eta_\mathrm{p}, \eta_\mathrm{s})=(5.5^\circ, 8^\circ)$.}
\label{fig: HOM Resonance scan SRC small etas}
\end{figure}

Figure\,\ref{fig: HOM Resonance scan SRC small etas} shows the same
HOM scan for $(\eta_\mathrm{p}, \eta_\mathrm{s})=(5.5^\circ, 8^\circ)$
with DRSE. The first HOM is not well suppressed, especially for
DRSE. This is a problem because the first order modes are strongly
coupled with mirror alignment fluctuations and easily excited. For
this reason, we did not employ the small Gouy-phase regions of
figure\,\ref{fig: HOM Resonance map PRC}.

\subsubsection{Gouy phase telescopes in the recycling cavities}
\label{sec:gouy-phase-telescope}

\begin{figure}[tbp]
\includegraphics[width=9cm]{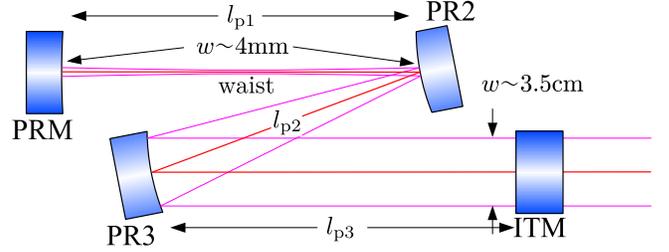}
  \caption{Schematic view of the folded power recycling cavity. The BS is omitted in this figure and the ITMs are combined into single effective ITM.}
  \label{fig:folding}
\end{figure}

In order to realize the desired Gouy phase shifts in the recycling
cavities, we have to focus the beams inside the cavities. We achieve
this by folding the cavities with two additional mirrors (folding
mirrors), to form a telescope. The schematic of the folding part of
the PRC is shown in figure\,\ref{fig:folding}. Although we mainly use the PRC
for the explanation in this section, the design for the SRC is almost
identical to the PRC. 

The total length of the cavity has to be equal to the one determined
in section\,\ref{sec:rf-sb-freq-len}. In order to reduce the
astigmatism, we want to separate the PR2 and the PR3 as far as
possible. With other practical constraints (mainly the size of the
vacuum chambers), we set the lengths of the folding part as shown in
table\,\ref{Table: folding params}.

After selecting the separations between the folding mirrors, we
optimize the ROCs of the folding mirrors. There are many combinations
of ROCs to realize a given Gouy phase shift in the PRC. The selected
values of the ROCs are shown in table\,\ref{Table: folding
  params}. Figure\,\ref{fig:folding} shows the beam profile of the PRC
with the selected ROCs. The beam from the ITMs is focused by the PR3
and hits the PR2 with a smaller beam size. The PR2 is a convex mirror
to collimate the beam. There is a beam waist at the middle of the PR2
and the PRM. Therefore, this design gives the same beam spot size on
the PRM and the PR2. Other combinations of the ROCs giving the same Gouy
phase change in the PRC make the beam spot sizes larger on one mirror and
smaller on the other compared with our design. A smaller spot size
increases a concern for thermal lensing. A larger beam makes it harder
to handle the beams at the REFL and the POP ports. The 4\,mm beam,
which is also well collimated, can be easily steered with standard 2
inch optics. The small beam size also makes it easy to separate
secondary reflection beams from, for example, the anti-reflection
surfaces of the ITMs. These stray beams have to be properly damped to
avoid scattered light noises. The large beam coming back from the arm
cavities (3.5cm radius) are not easy to separate. We utilize the beam
reducing functionality of the Gouy phase telescope as an extra benefit
to cleanly separate the stray beams between the PR2 and the PRM after
the beam sizes are reduced.

Although the desired values of the Gouy phase shifts in the PRC and
the SRC are slightly different, we decided to use almost the same ROCs
for the folding mirrors of the two cavities, so that the same
reference spheres can be used for polishing. We slightly changed the
length parameters of the SRC from the PRC to realize the different
Gouy phase shift with only a minimal change of the ROC of the SR2. The
total length of the SRC is not changed with this adjustment.

Table\,\ref{Table: PRC thermal lens} shows expected thermal lens
effects on the PRC mirrors. The following formula is used to asses the
effective change of the ROC by thermal
lensing\,\cite{winkler_heating_1991}.
 
\begin{equation}
  \label{eq:3}
  dR = \frac{\alpha R^2 P_\mathrm{a}}{2\pi\kappa w^2},
\end{equation}
where, $R$ is the ROC of the mirror, $\alpha$ is the thermal expansion
coefficient of the substrate, $\kappa$ is the thermal conductivity,
$w$ is the beam spot radius on the mirror and $P_\mathrm{a}$ is the
absorbed light power at the surface of the mirror.  We assumed an
intra-cavity power of 800\,W and 10\,ppm absorption loss at the
reflection of each mirror. Although the 10\,ppm absorption is rather
large, we take it as a safety margin. The amount of the ROC change
from the thermal lensing is less than the figure error tolerance
discussed in the next section.

\begin{table}
\caption{Parameters of the folding cavities. $l_\mathrm{p1}$, $l_\mathrm{p2}$ and $l_\mathrm{p3}$ are the lengths of the three segments of the folded PRC as shown in figure\,\ref{fig:folding}. $l_\mathrm{s1}$, $l_\mathrm{s2}$ and $l_\mathrm{s3}$ are the corresponding lengths of the SRC. \label{Table: folding params}}
\begin{ruledtabular}
\begin{tabular}{cc|cc}
$l_\mathrm{p1}$ &14.762\,m&PRM ROC&458.129\,m\\
$l_\mathrm{p2}$ &11.066\,m&PR2 ROC&-3.076\,m\\
$l_\mathrm{p3}$ &15.764\,m&PR3 ROC&24.917\,m\\
$l_\mathrm{s1}$ &14.741\,m&SRM ROC&458.129\,m\\
$l_\mathrm{s2}$ &11.112\,m&SR2 ROC&-2.987\,m\\
$l_\mathrm{s3}$ &15.739\,m&SR3 ROC&24.917\,m\\

\end{tabular}
\end{ruledtabular}
\end{table}

\begin{table}
\caption{Thermal lens effect on the PRC mirrors}
\label{Table: PRC thermal lens}
\begin{ruledtabular}
\begin{tabular}{llll}
Mirror&Beam radius&$dR$&Tolerance\\\hline
PRM &4\,mm&5.5\,m&$\pm 20$\,m\\
PR2 &4\,mm&0.24\,mm&$\pm 10$\,mm\\
PR3 &35\,mm&0.21\,mm&$\pm 10$\,mm\\
\end{tabular}
\end{ruledtabular}
\end{table}

\subsubsection{ROC error}
\label{sec:roc-error-recovery}

The mode profile of the Gouy phase telescope is highly sensitive to the
errors in the ROCs of the mirrors, especially of the PR2 and the PR3. If the
mode of the PRC is not matched with the arm cavity modes, the
recycling gain is reduced. In the case of mode mismatch between the SRC and the
arm cavities, the gravitational wave sidebands coming out to the AS port are
reduced. In addition, a mode profile change is usually associated with
deviation of the Gouy phase shift from the desired value.

\begin{figure}[tbp]
\includegraphics[width=8cm]{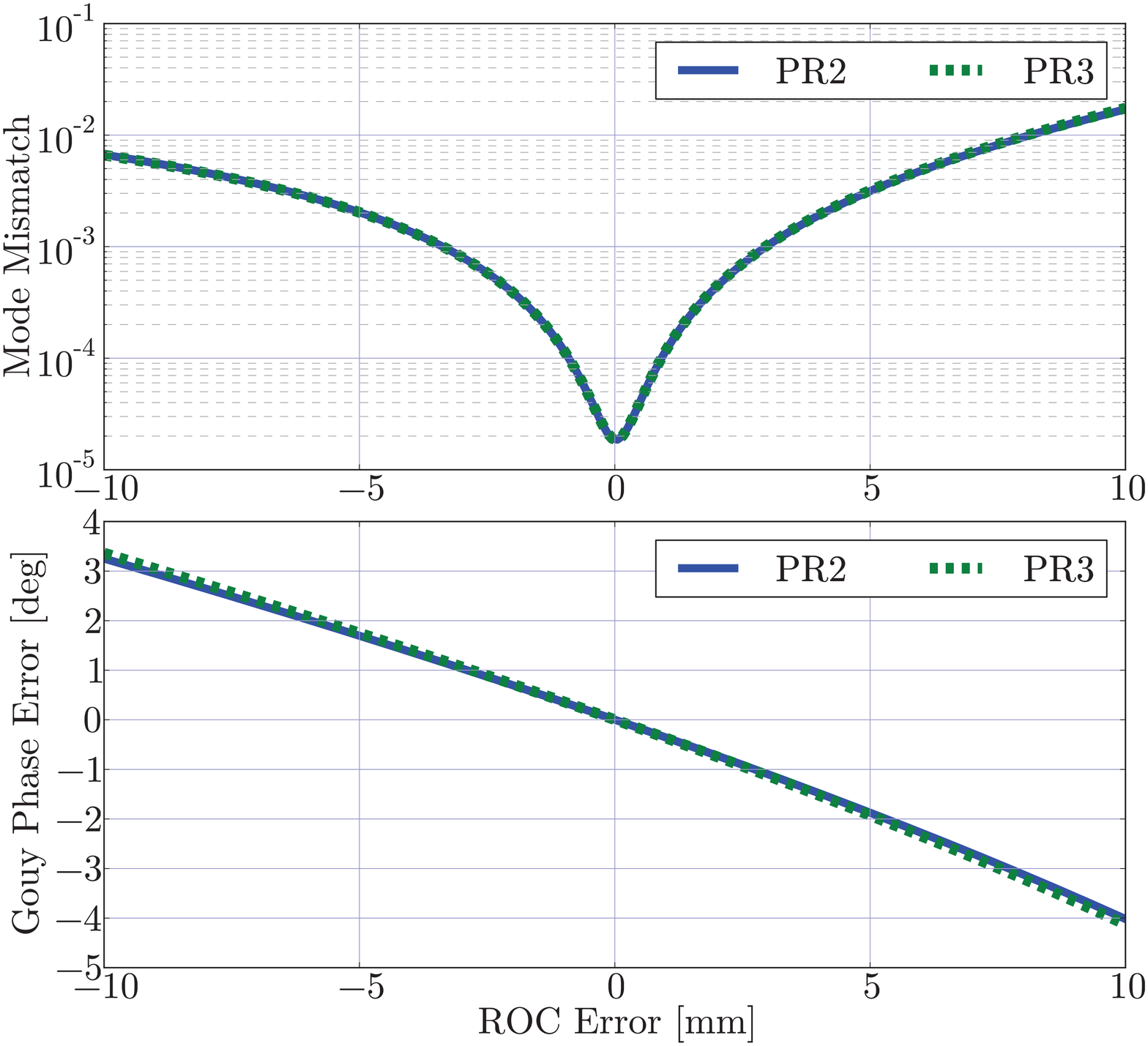}
  \caption{The mode mismatch and the Gouy phase error as functions of
    the ROC errors of the PR2 and the PR3.}
  \label{fig:ROC-error-Mode-mismatch}
\end{figure}

Figure\,\ref{fig:ROC-error-Mode-mismatch} shows the mode-mismatching value
and the deviation of the one-way Gouy phase shift from the desired value,
plotted against the error in the ROCs of PR2 and PR3. The
mode-mismatching value is defined by the following formula:
\begin{equation}
  \label{eq:4}
  1 -  \left|\int^\infty_{-\infty}\int^\infty_{-\infty}\psi^*_\mathrm{PRC}(x,y)\psi_\mathrm{ARM}(x,y) dxdy\right|^2,
\end{equation}
where $\psi_\mathrm{PRC}(x,y)$ and $\psi_\mathrm{ARM}(x,y)$ are the
complex beam profile functions of the PRC and the arm cavity modes,
representing the electric field amplitude and the phase of the laser beams
in a cross sectional plane. These are normalized to make the integral
1 when the two modes are identical.

The plots show that if we want to keep the mode matching above 99\%,
the ROC error of the PR2 and the PR3 have to be less than 5\,mm. The 5\,mm ROC
error also gives a Gouy-phase deviation of about $4^\circ$. 

\begin{figure}[tbp]
\includegraphics[width=8cm]{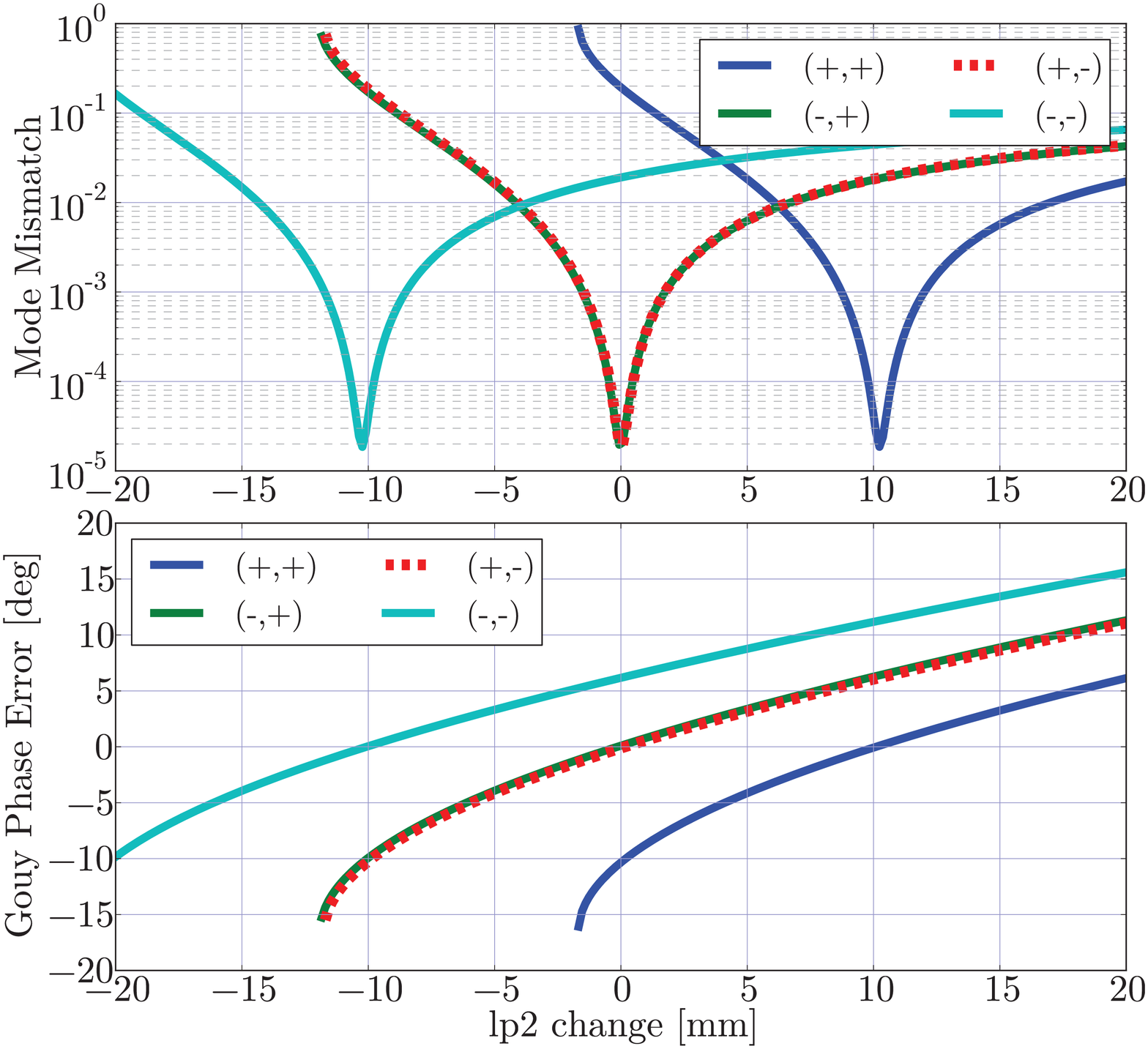}
  \caption{The mode mismatch and the Gouy phase error as functions of
    $l_\mathrm{p2}$. The curves show when ROC errors of 10\,mm is added to the PR2 and
    the PR3. The signs of the introduced errors are indicated in the
    legend.}
  \label{fig:Lp2-Mode-mismatch}

\end{figure}

Although keeping the ROC error to be less than 5\,mm out of the
24\,m ROC of the PR3 is not easy to achieve, we can recover the error
by changing the distance between the PR2 and the PR3
($l_\mathrm{p2}$). Figure\,\ref{fig:Lp2-Mode-mismatch} shows how the
mode matching and the Gouy-phase error change when $l_\mathrm{p2}$ is changed with
ROC errors of 1\,cm added to the PR2 and the PR3. The curves on the plot
show the four possible combinations of the signs of the errors on the
two mirrors. Since we have to keep the total PRC length constant, $l_\mathrm{p1}$
is changed at every point to compensate for the change in $l_\mathrm{p2}$.

Even with the worst combination of the errors, the mode-matching can
be recovered by changing $l_\mathrm{p2}$ by roughly the same amount as
the ROC error. The one-way Gouy phase is also recovered to the desired
value with the same adjustment of $l_\mathrm{p2}$. When the mirrors
are installed, we plan to adjust $l_\mathrm{p2}$, based on the measured
value of the ROCs of the fabricated PR2 and PR3. From this
observation, the tolerances for the mirror polishing error of the PR2
and PR3, shown in table\,\ref{Table: PRC thermal lens}, are set to the
value by which the suspension systems can be moved without too much hustle. The
effect of the ROC error of the PRM on the mode matching is
moderate. We set its error tolerance by requiring the mode matching be
greater than 99.99\%.

\section{Alignment signals}
\label{Section: Alignment}

\subsection{Wave-front sensor shot noise}
\label{sec:wave-front-sensor}

With the interferometer parameters selected in the preceding sections,
we have to check whether reasonable error signals for the alignment
control can be obtained. For the alignment signal extraction, we use
the wave-front sensing (WFS)
technique\,\cite{morrison_automatic_1994}. Two quadrant photo
detectors (QPDs) are placed at each detection port of the
interferometer to receive the output beam at two orthogonal Gouy
phases. Signals from each quadrant of the QPDs are demodulated either
at the f1 or the f2 sideband frequencies. The difference between the
demodulated signals from the left quadrants and the right quadrants
yields an error signal proportional to the rotations of the mirrors in
the horizontal plane (Yaw). The difference between the upper and lower
quadrants yields an error signal for the bowing rotations of the
mirrors (Pitch).

In order to compute the WFS signals for a given interferometer
configuration, we again used Optickle. By calculating the transfer
functions from the rotations of the mirrors to QPD signals at
various ports, one can obtain a sensing matrix of dimensions
(number of signal ports) $\times$ (number of mirrors). Since the
sensing matrix is not diagonal in general, we take a linear
combination of the QPD signals $q_n$, such that $\theta_m = \sum_n a_n
q_n$, to extract a diagonalized rotation signal $\theta_m$ of the $m$-th mirror,
where $n$ distinguishes QPDs and $a_n$ is the weighting factor.
Optickle can also compute the shot noise $s_n$ of each QPD output. The
total shot noise of each diagonalized signal is calculated by
\begin{equation}
  \label{eq:WFS shot noise}
\theta^\mathrm{shot}_m = \sqrt{\sum_n(a_n s_n)^2}.  
\end{equation}
Note that even if the shot noise of each QPD output is small, the
diagonalized shot noise can be large, if the mirror rotation signals
are nearly degenerated in the QPD signals. Therefore, (\ref{eq:WFS
  shot noise}) can also be used as a figure of merit for good signal
separation.

\subsection{Coupling of WFS noises to DARM}
\label{sec:coupling-wfs-noises}

Once the shot noises of the WFS signals are calculated, we now
consider their contributions to the bottom-line sensitivity. The
rotation $\theta_m$ of a mirror is converted to the length change
$\delta L_m$ of an optical path, coupled with the beam mis-centering
$d_m$ from the rotational center of the mirror:
\begin{equation} \label{Eq: A2L}
 \delta L_m = d_m\theta_m.
\end{equation}
For an alignment servo loop with an open loop gain of $G_m(f)$, the WFS shot noise $\theta^\mathrm{shot}_m$ is converted to the actual rotation $\theta_m$ of the mirror by
\begin{equation} \label{Eq: shotnoise induced rotation}
 \theta_m(f) = \frac{G_m(f)}{1+G_m(f)} \theta_m^{\rm shot}.
\end{equation}
The transfer function $k_m(f)$ from the motion $\delta L_m$ of the
$m$-th mirror to the DARM signal can be computed with the Optickle
model described in section\,\ref{Section: RFSB} (see also
\cite{aso_length_2012}). By requiring $k_m(f)\delta L_m(f)$ to be
smaller than the target sensitivity $h(f)$ in the observation frequency band
(above 10\,Hz), we can derive requirements for the WFS shot noise
levels as,
\begin{equation}
\label{eq: shot noise req}
 \theta_m^{\rm shot} < \mathrm{Min}\left[\frac{h(f)}{d_m k_m(f)} \frac{1+G_m(f)}{G_m(f)},  f>10\mathrm{Hz}\right].
\end{equation}

The beam mis-centering $d_m$ depends on two factors. One is the static
mis-centering, which is how well we can adjust a beam spot position at
the center of a mirror. From the experience of the first generation
detectors, we assume the accuracy of this adjustment to be 0.1\,mm.
Secondly, the alignment fluctuations of the interferometer mirrors
cause the beam spots to move around. The conversion coefficient from
the rotation angle of a mirror to the beam spot position changes in
the other mirrors can be calculated with Optickle. Assuming RMS
angular fluctuations of the mirrors to be less than $10^{-8}$\,rad,
RMS beam spot motions on the mirrors are estimated to be smaller than
0.1\,mm. Therefore, we use $d_m$=0.1\,mm for the test masses in the
following calculations. For the other mirrors, we relax the requirement
and assume $d_m$=1\,mm.

The assumed angular fluctuation RMS of $10^{-8}$\,rad comes from the
requirement for the beam jitter coupling, calculated for KAGRA using
the method described in \,\cite{mueller_beam_2005}. It has to be
ensured by the local damping of the suspension systems in combination
with the WFS servo. Detailed analysis of this requires an elaborate
suspension model and it will be reported elsewhere. We only make a
quick comment here that the RMS angular fluctuations of the KAGRA
suspensions are mostly determined by rotational resonances below
0.5Hz. Therefore, it is likely that we can suppress those resonances
by WFS servos with UGFs of 1\,Hz or above, which are assumed in the
following analysis.

\begin{table}
    \caption{WFS shot noise requirements and the simulated shot noises. All values are in the unit of ${\rm rad/\sqrt{Hz}}$. \label{Table: WFSshot}}
\begin{ruledtabular}
\begin{tabular}{ccccc}
 & \multicolumn{2}{c}{BRSE} & \multicolumn{2}{c}{DRSE}  \\
 & Requirement & Simulated & Requirement & Simulated  \\
      \hline
ETMX & $8.8\times 10^{-15}$ & $1.9\times 10^{-14}$ & $9.7\times 10^{-15}$ & $2.9\times 10^{-14}$ \\
ETMY & $8.8\times 10^{-15}$ & $1.9\times 10^{-14}$ & $9.7\times 10^{-15}$ & $1.9\times 10^{-14}$ \\
ITMX & $8.8\times 10^{-15}$ & $2.8\times 10^{-14}$ & $9.7\times 10^{-15}$ & $3.7\times 10^{-14}$ \\
ITMY & $8.8\times 10^{-15}$ & $2.8\times 10^{-14}$ & $9.7\times 10^{-15}$ & $2.8\times 10^{-14}$ \\
BS   & $9.2\times 10^{-12}$ & $7.4\times 10^{-13}$ & $1.5\times 10^{-11}$ & $3.1\times 10^{-12}$ \\
PR3  & $3.2\times 10^{-09}$ & $2.7\times 10^{-13}$ & $1.4\times 10^{-09}$ & $1.1\times 10^{-12}$ \\
PR2  & $3.2\times 10^{-09}$ & $1.0\times 10^{-13}$ & $1.4\times 10^{-09}$ & $3.1\times 10^{-13}$ \\
PRM  & $3.2\times 10^{-09}$ & $8.9\times 10^{-14}$ & $1.4\times 10^{-09}$ & $6.1\times 10^{-13}$ \\
SR3  & $7.4\times 10^{-12}$ & $7.7\times 10^{-12}$ & $1.3\times 10^{-11}$ & $1.3\times 10^{-11}$ \\
SR2  & $7.4\times 10^{-12}$ & $6.6\times 10^{-11}$ & $1.3\times 10^{-11}$ & $1.2\times 10^{-10}$ \\
SRM  & $7.4\times 10^{-12}$ & $1.4\times 10^{-12}$ & $1.3\times 10^{-11}$ & $6.8\times 10^{-12}$ \\
\end{tabular}
\end{ruledtabular}
\end{table}

Table\,\ref{Table: WFSshot} summarizes WFS shot noise requirements and
calculated shot noises for our interferometer configuration.  For the
calculation of the requirements, we assumed the UGFs of $G_m(f)$ to be
3\,Hz for the test masses and 1\,Hz for the other mirrors. The shape
of $G_m(f)$ is $1/f$ around the UGF and $1/f^4$ cut-off is added at
10\,Hz for the test masses and at 3\,Hz for the other mirrors. Because
the effective resonant frequency of the negative angular optical
spring of the arm cavities (section\,\ref{sec:g-factor}) is at about
1\,Hz, the WFS servos of the test masses have to suppress this
instability. Therefore, the UGF of the test mass servos is set
higher. Each requirement value in the table is computed with (\ref{eq: shot noise
  req}) and then divided by $\sqrt{22}$ to take into account the fact
that there are incoherent noise contributions from 11 mirrors with 2
rotational DOFs each.

\begin{figure*}[tbp]
  \begin{minipage}{8cm}
\includegraphics[width=8cm]{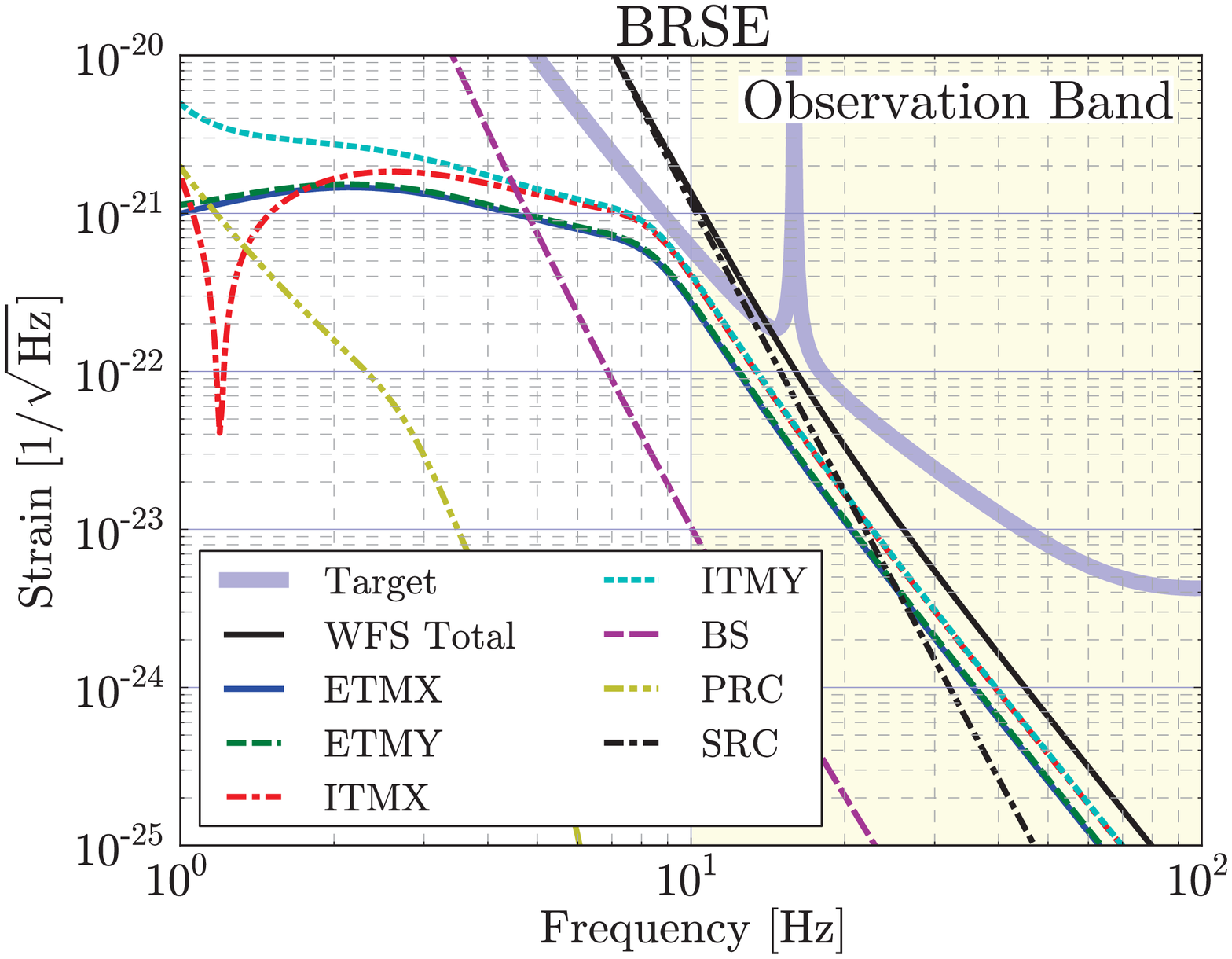}    
  \end{minipage}
\hspace{0.5cm}
  \begin{minipage}{8cm}
\includegraphics[width=8cm]{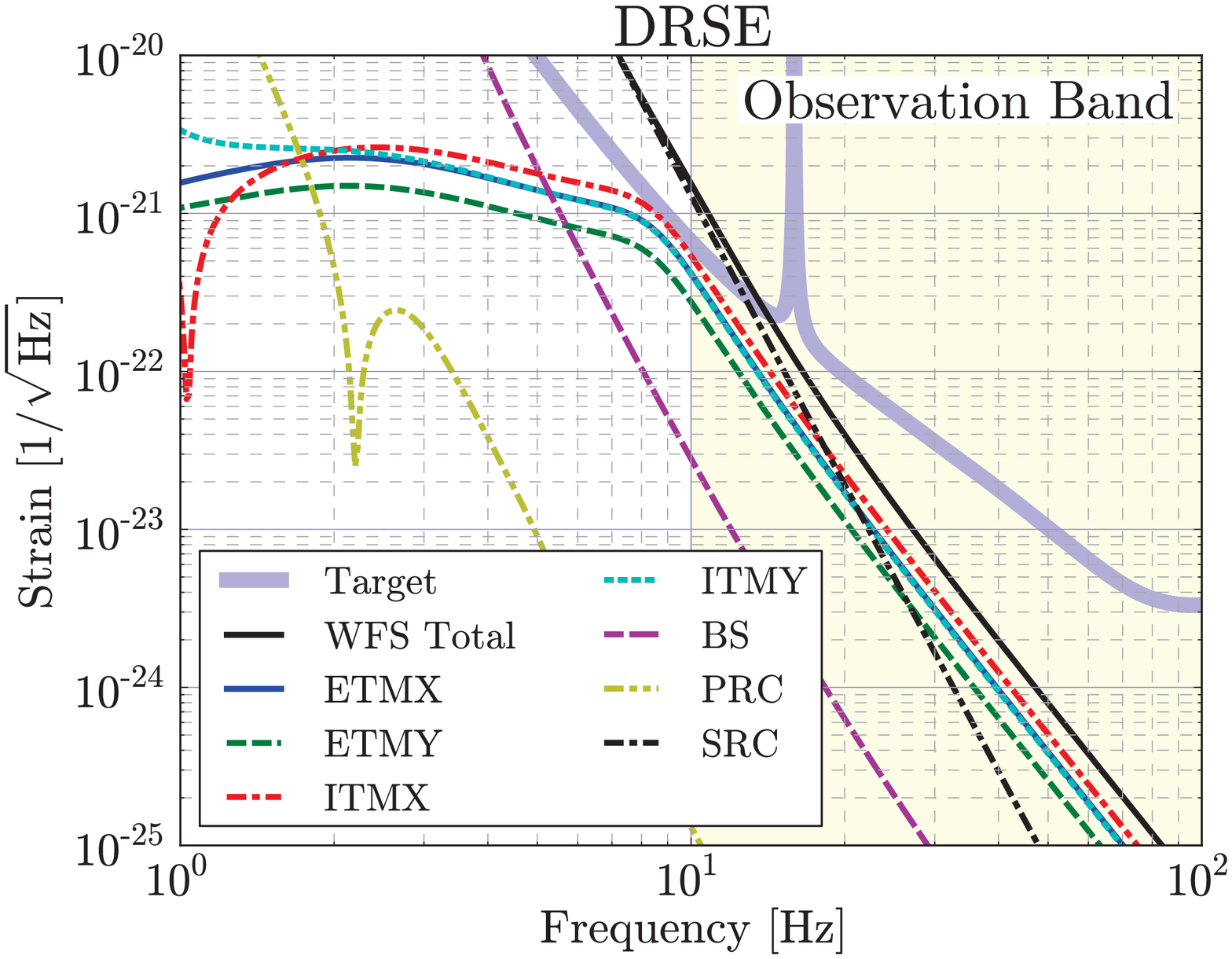}    
 \end{minipage}
\caption{Contributions of WFS shot noises to the strain sensitivity. The PRC and the SRC consist of three mirrors each. Therefore, the curves labeled PRC and SRC are the quadratic sums of the noises from the three mirrors.}
\label{fig:WFS Spectra}
\end{figure*}

\begin{figure*}[tbp]
  \begin{minipage}{8cm}
\includegraphics[width=8cm]{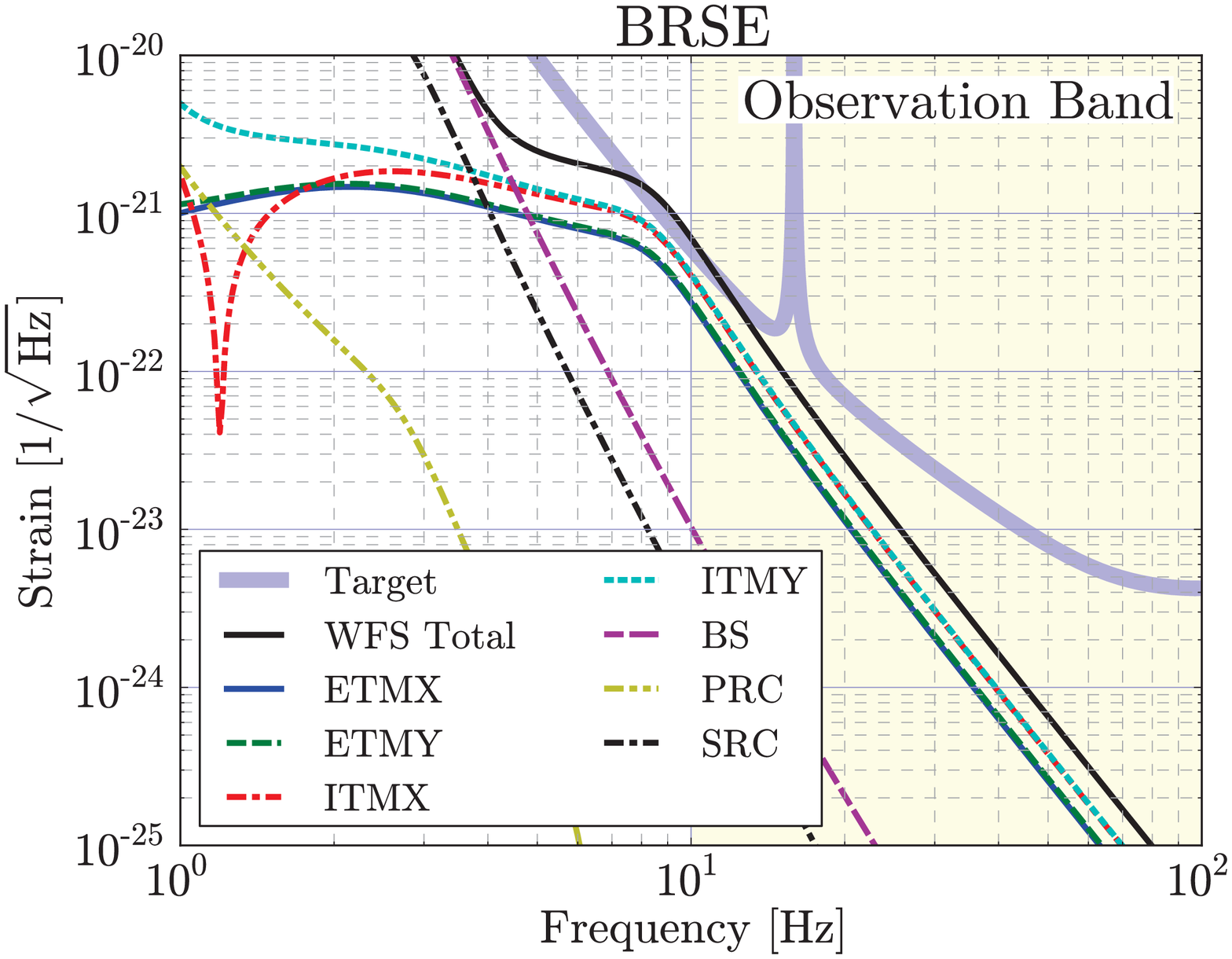}    
  \end{minipage}
\hspace{0.5cm}
  \begin{minipage}{8cm}
\includegraphics[width=8cm]{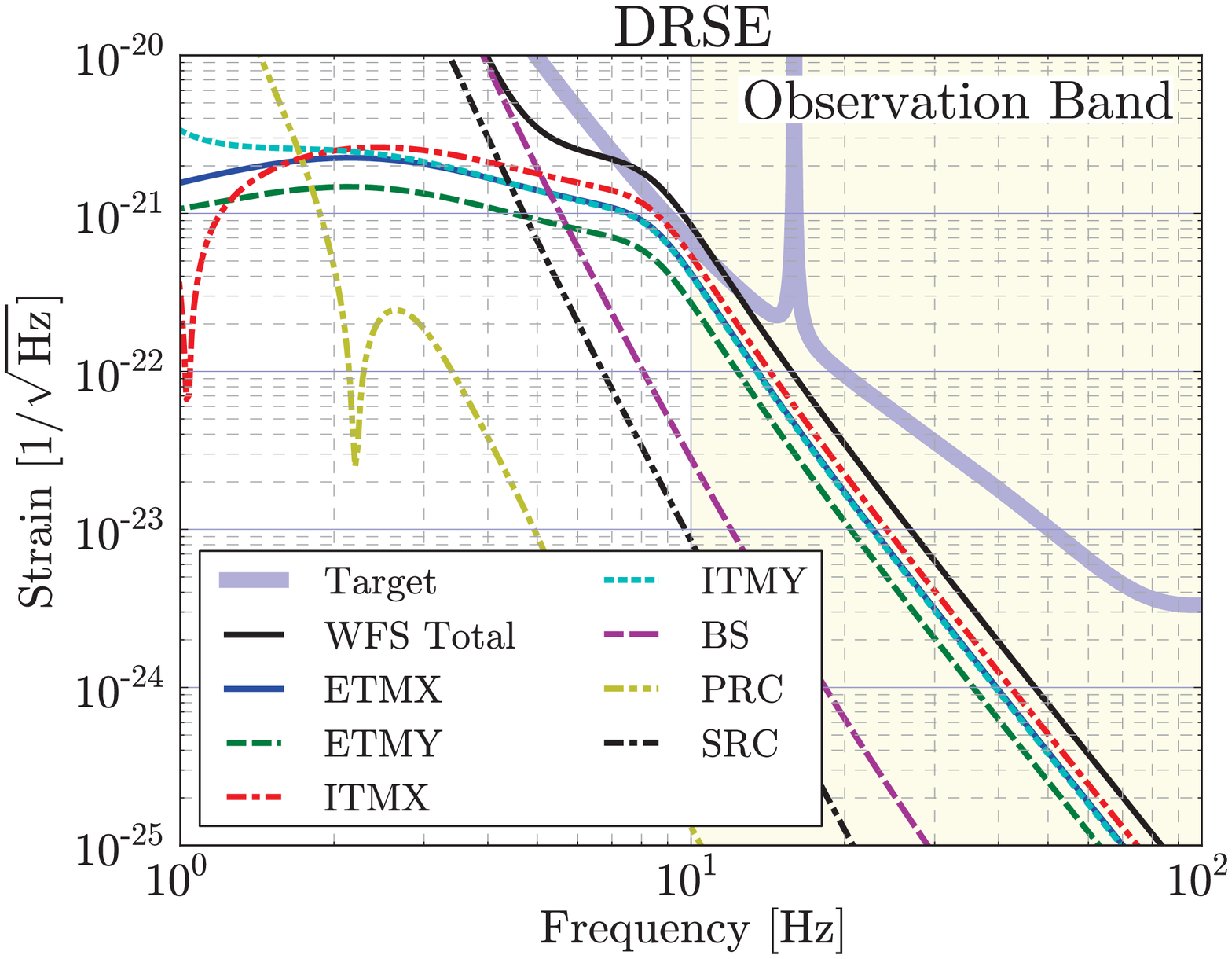}    
 \end{minipage}
\caption{Contributions of WFS shot noises to the strain sensitivity when the SR2 is not controlled by WFS.}
\label{fig:WFS Spectra No SR2}
\end{figure*}

The test masses have 2 to 4 times larger shot noises than the
requirements. SR3 and SR2 also do not satisfy the shot noise
requirements.  Figure\,\ref{fig:WFS Spectra} shows the spectral
contributions of the WFS shot noises to the strain sensitivity. At the
lower edge of the observation band (around 10\,Hz), the total WFS shot
noise exceeds the target sensitivity. Especially, the SRC contribution
is large because of the poor shot noise of the SR2 signal. In the
actual operation, we may not control the SR2 using WFS or use it only
as a DC reference for alignment. Figure\,\ref{fig:WFS Spectra No SR2}
shows the WFS shot noises when SR2 is not controlled. In this case,
the SRC contribution is reduced significantly. However, the WFS shot
noises of the test masses still touch the target sensitivity at
10\,Hz.

Although the noise requirements are not strictly satisfied, since the
noise excess happens only at the very edge of the observation band,
where the sensitivity is not so good in the first place, the impact on
the IR is minimum. The calculated IR reductions by the WFS noise are
less than 1Mpc for all the cases. A more serious concern is that in
an actual operation, the WFS noises may be higher than the estimated shot
noises for various technical reasons. We do not have any safety margin
for the test masses to tolerate such a noise increase. In this case,
we have to further reduce the beam mis-centering, decrease the UGF or
use more aggressive cut-off filters.



\section{Conclusion}
\label{Section: Conclusion}
We explained the detailed design process of the KAGRA interferometer,
starting from classical noises as boundary conditions. Then we
optimized the quantum noise shape using binary inspiral ranges as
guidance. Scientific and risk related reasoning led us to decide to
make KAGRA capable of operating in both BRSE and DRSE
configurations. We then developed a length sensing scheme using two sets of RF
modulation sidebands. The ROCs of the mirrors were chosen to make the
interferometer robust against unwanted higher order mode resonances.
Alignment sensing noise couplings were examined to asses the impact on
the target sensitivities.

The most serious concern in the current design is the nearly zero
noise margin for the alignment control. Reduction of the alignment
control UGFs can reduce the noise couplings significantly. Whether
this is possible or not depends on the detailed control design of the
suspension systems and their local damping systems. Works on this
issue with both computer simulations and experimental verification are
on going.

With the above considerations, the parameters of the KAGRA
interferometer are fully determined. Fabrications of the components
are now underway and the installations will start in 2014.

\begin{acknowledgments}
  The authors are grateful to Rana Adhikari, Koji Arai, Keita Kawabe,
  Matt Evans, Lisa Barsotti and Mike Smith for helpful discussions and
  advice. This work was supported by the Leading-edge Research
  Infrastructure Program of Japan. The LIGO Observatories were
  constructed by the California Institute of Technology and
  Massachusetts Institute of Technology with funding from the NSF
  under cooperative agreement PHY-9210038. The LIGO Laboratory
  operates under cooperative agreement PHY-0107417.

\end{acknowledgments}

\bibliography{KAGRA_MIF,KAGRA_MIF_Misc}

\end{document}